\documentclass[letterpaper,twocolumn,10pt]{article}
\usepackage{templates/usenix2019,epsfig,endnotes}

\usepackage{times}
\usepackage{algpseudocode}
\usepackage{algorithm}

\usepackage{algorithm, algorithmicx, algpseudocode}

\usepackage{url}
\usepackage{booktabs} 
\usepackage{balance}
\usepackage{float}
\usepackage{subfig}
\usepackage{graphicx}
\usepackage{listings}
\usepackage{setspace}
\usepackage{xspace}
\usepackage{color}

\usepackage{tikz}

\usepackage{enumitem}

\usepackage[font={footnotesize,it}]{caption}
\usepackage[export]{adjustbox}
\usepackage[skip=1ex]{caption}

\usepackage[small, compact]{titlesec}
\titlespacing*{\section}{1pt}{3.5pt}{2pt}
\titlespacing*{\subsection}{1pt}{3pt}{1.5pt}
\titlespacing*{\subsubsection}{1pt}{3pt}{1.5pt}

\usepackage{amssymb}
\usepackage{pifont}

\newcommand{\secref}[1]{{\S\ref{#1}}}

\newcommand{\sujata}[1]{}
\newcommand{\arjun}[1]{}
\newcommand{\junaid}[1]{}
\newcommand{\aditya}[1]{}
\renewcommand{\sujata}[1]{{\color{red}{\bf SB:#1}}}
\renewcommand{\arjun}[1]{{\color{brown}{\bf AS:#1}}}
\renewcommand{\aditya}[1]{{\color{blue}{\bf AA:#1}}}
\renewcommand{\junaid}[1]{{\color{orange}{\bf JK:#1}}}

\newcommand{\myvec}[1]{\protect\overrightarrow{#1}}

\newcommand{\compactcaption}[1]{\vspace{-0.3em}\caption{#1}\vspace{-1.5em}}

 \newenvironment{denseitemize}{
 \begin{itemize}[noitemsep, topsep=0pt, leftmargin=1.2em]
 }{\end{itemize}}

\newcommand{\name}{SNF\xspace}

\definecolor{nodecolor}{RGB}{255,115,115}

\begin{document}

\title{\name: Serverless Network Functions}
\linespread{1}
\author{\\ \rm Arjun Singhvi$^{*}$, Junaid Khalid$^{*}$, Aditya Akella$^{*}$, Sujata Banerjee$^{\dag}$\\
		 \rm University of Wisconsin - Madison$^{*}$ \hspace{0.6em} VMware Research$^{\dag}$
	    }

\date{}
\maketitle

	\noindent

{\bf{\em Abstract---}} It is increasingly common to outsource network
functions (NFs) to the cloud. However, no cloud providers offer
NFs-as-a-Service (NFaaS) that allows users to run custom NFs. Our work
addresses how a cloud provider can offer NFaaS. We use the emerging
serverless computing paradigm as it has the right building blocks -
usage-based billing, convenient event-driven programming model and
automatic compute elasticity. Towards this end, we identify two core
limitations of existing serverless platforms to support demanding
stateful NFs - coupling of the billing and work assignment
granularities, and state sharing via an external store. We develop a
novel NFaaS framework, \name, that overcomes these issues using two
ideas. \name allocates work at the granularity of {\em flowlets}
observed in network traffic, whereas billing and programming occur on
the basis of packets. \name embellishes serverless platforms with {\em
ephemeral state} that lasts for the duration of the flowlet and
supports high performance state operations between compute units in a
peer-to-peer manner. We demonstrate that our \name prototype
dynamically adapts compute resources for various stateful NFs based on
traffic demand at very fine time scales, with minimal overheads.

	\section{Introduction}
\label{sec:introduction}

Infrastructure- and compute-as-a-service relieves cloud users from the
burden of managing physical compute resources and scaling capacity. A
recent exciting advancement in the cloud evolution is {\em serverless
  computing}, which further reduces the burden of managing,
provisioning, and scaling infrastructure (servers, VMs, etc.).  Now all
a user has to focus on is the application logic, which tremendously
reduces the barrier to entry.  In this model,
users write and upload programs called ``functions'' that can be
dynamically scaled up and down based on user specified event
triggers. The functions run in stateless and short-lived
computing instances/containers, and the user is billed for exactly the
computing cycles used for a function, providing significant cost
advantages.  While this is a big step in the right direction towards
offering services that are cost-effective for users, serverless
computing today is rather narrowly focused. It targets stateless,
short-lived, batch mode, and/or embarrassingly parallelizable
workloads.  It leaves behind stateful streaming workloads, which form
our focus.

Rather than consider general stateful streaming applications, we
consider the specific and important example of network functions (NFs)
that perform stateful operations on packet streams in enterprises and
telecommunication infrastructures.  These workloads in recent years
have emerged as part of the Network Functions Virtualization (NFV)
transformation in telco service provider networks and still face
significant challenges in simultaneously achieving the goals of low
cost (to both the user and the infrastructure provider), and scalable
performance, while supporting programming ease. In many ways,
serverless computing provides a key set of building blocks to address
these issues impacting NFV, but important gaps remain
(\secref{sec:motivation}).

Our high-level goal is  to address the challenges of
supporting NF workloads with serverless computing infrastructures by
adding in the missing abstractions and mechanisms to support stateful
computation effectively. Also, we should not have to compromise on
preserving the unique benefits of the serverless paradigm, such as
simplified management, event-driven compute triggering which is
central to usage-based pricing, autoscaling, and
cost-effectiveness. This is in contrast to prior works that {\em
  retrofitted} applications onto today's serverless platforms~\cite{ex-camera,pywren}. 

Network functions are a resource intensive workload with fine grained
latency and throughput performance needs. We examine limitations
imposed by two naive realizations of support for such workloads over
today's serverless platforms---invoking a function to process each
individual {\em packet}, or invoking a function instance per {\em
  flow}. 
We make two key observations, and argue,
correspondingly, for two key design changes. First, we observe that,
in today's platforms, ``events'' (e.g., each incoming request) are
used for both the granularity of work allocation as well as the
granularity against which functions are programmed and billing is
performed. This coupling imposes a hard trade-off between resource use
efficiency, performance, and billing granularity for stateful
applications. We argue for {\em breaking this coupling}, and allowing
work allocation to happen at a different granularity than that at
which a program launched in a function operates. Second, we observe
that given state externalization in today's platforms (externalization
is key to keeping functions stateless, enabling rapid scaleout) the
only way state can be shared across events and functions is by using a
remote storage service. State sharing is crucial to stateful
applications and, for NFs, externalization substantially worsens
packet processing latency. Thus, we advocate for {\em ephemeral
  statefulness}, where state is bound to function instances for the
duration of the computation corresponding to a single unit of work
allocation.

We leverage these two ideas in designing, \name, a new serverless platform that allows cloud providers to provide NFs-as-a-service (NFaaS) wherein users can outsource NFs to enjoy the benefits of the cloud~\cite{aplomb, embark}. Users of \name (e.g.,
NFV operators) can program many different NFs as functions. For a
given NF, \name transparently distributes the packet processing work
in an incoming traffic stream across an elastically scalable set of
compute units; each unit has a function corresponding to the given NF
deployed in it. \name maintains high utilization of compute units, and
ensures that minimal number of units are used, both of which are
attractive to the cloud provider. \name also ensures that a given
NF deployment's packet processing throughput is high, tail latency is
low, and that billing only captures the work done processing traffic,
all of which are appealing to users (NFV operators).  

To achieve these, \name relies on the following ideas:
\begin{denseitemize}
\item We note that the packet processing workloads in a flow can be
  naturally granularized into {\em flowlets}~\cite{flowlets}. In
  \name, we use flowlets as the units of our workload assignment,
  whereas NF programs that run in function instances operate on a
  packet at a time, preserving the NF programming abstraction today
  and ensuring billing only for work done.

\item We store NF-internal state in the local memory of a compute unit
  where a given flowlet is being processed.  We develop protocols that
  uses inter flowlet gaps to proactively replicate ephemeral state to
  a new compute unit where the processing of a subsequent flowlet in
  the same flow is to occur, while avoiding inconsistent updates.

\item To achieve efficiency, our work assignment algorithm aims
    to keep all active compute units at the maximum possible
    utilization via a novel weighted greedy bin-packing algorithm that
    maximally packs flowlets into few compute units, while ensuring
    performance targets are met, and while preferring instances to
    which state has been proactively replicated.
    
\end{denseitemize}

We implement and evaluate a standalone prototype of \name and deploy
on CloudLab~\cite{cloudlab}. Our experimental results
with real traffic traces and five stateful NFs demonstrate the
viability of our architecture and validate the core decisions to use
flowlets and ephemeral state. With \name, we are able to
simultaneously achieve efficiency, performance and fault tolerance for
NF processing. We are able to closely match the packet processing demand and provisioned resources dynamically at fine grained timescales of 100ms, whereas naive NF implementations over serverless architectures result in both over and underload. Our results show that \name reduces 75\%-ile processing latency by \textbf{2.9K-19.6Kx} over alternatives that operate at flow granularity. Our proactive state management improves the 99\%-ile tail latency of NF processing by \textbf{12-15x} over state-of-the-art state management solutions. Additionally, our simple fault tolerance protocol supports fast recovery (\textbf{22.8x-183.6x} reduction in comparison to alternatives).

	\section{Motivation}
\label{sec:motivation}

\subsection{Why Serverless Computing for NF-as-a-Service?} 
 
NFs examine and modify packets to ensure security, improve performance, and
provide other diverse functionalities: examples
include network address translators (NAT), intrusion detection systems (IDS),
firewalls, load balancers, etc. Many NFs are stateful in nature, for example, NAT
maintains address mappings, IDS maintains string matches , etc. Over the
last few years, researchers have advocated outsourcing NFs to the
cloud~\cite{aplomb, embark}. These works have observed that such outsourcing
to realize NFaaS can enable NF users to enjoy the benefits provided by the cloud such as
leveraging the scale, elasticity, and availability of the cloud, pay-as-you-go
billing that is intrinsic to the cloud, and the built-in management and
operational expertise available at cloud providers.

However, as of today, no cloud provider offers NFaaS. There
exists limited support wherein an end user can use specific out-of-the-box NFs
provided by the cloud provider, such as a load balancer, or a firewall~\cite{aws-lb, aws-firewall}. 
But, the user does not have the flexibility of
running custom NFs. 
Our work addresses how a cloud provider can offer NFaaS and realize the goal of outsourcing NFs.

Ideally, an NFaaS
platform should provide an intuitive programming model that allows users to
write custom NF logic and delivers good performance (low packet processing
latency) while automatically managing the infrastructure (scaling up/down) to
meet the traffic demand and charging users only for the work
performed, i.e., usage-based billing.

Recently, a new cloud computing paradigm known as function-as-a-service,
or  serverless computing, has garnered a lot of attention,
primarily due to its attractive features such as event-driven programming
model, usage based billing, and automatic compute elasticity. Additionally,
serverless computing with fine-grained functions  is also beneficial to the cloud provider as it can
facilitate higher resource utilization, both in terms of short
execution time and small resource footprint in comparison to other compute
alternatives such as VMs. Thus, on the face of it, it appears that
serverless computing has the right building blocks to meet the aforementioned
requirements of NFaaS, which a cloud provider could leverage.

\subsection{Drawbacks of other realizations of NFaaS} 

It is possible for a provider
to use VM- or container-based compute platforms to realize NFaaS. However,
there are a few  fundamental impediments that arise. First, today’s native VM-
or container-based compute platforms’ interface does not allow users to simply supply
high-level functions; users are responsible for managing the lifecycle of compute units (e.g., launch the compute unit, install appropriate NF logic and other software dependencies etc) which imposes significant
management burden.\footnote{It is possible to provide abstractions that allow
users to simply supply functions, but this is precisely what serverless
platforms already support.} Second, existing compute platforms charge on
the basis of the amount of time that a compute unit is assigned to the user. This used to
be in hourly increments, but has more recently been made more fine-grained.
Despite this improvement, pay-as-you-go charging today is fundamentally not
tied to actual usage -- in true usage-based pricing, the user is charged only
when the compute unit is actively processing user data or running user’s
compute logic. This undermines the cost effective reasons for users wanting to leverage NFaaS. Third, VM and
container-based compute platforms were designed with the notion of
virtualizing entire servers, with the aim of eliminating physical server
deployment/management and to support consolidation. Because of this, the
platforms are most useful in supporting batch style stateful, server-like computations.
However, similar to applications such as database services and web services, the NFaaS workload can be bursty, and consequently the cloud provider would suffer from low resource utilization as resources are typically
provisioned for peak load~\cite{berkeley-serverless-report}.

\subsection{Issues with Serverless Computing} 

Given that the goal of outsourcing NFs
aligns with the features of
serverless computing, we now briefly provide an overview of the paradigm and
then go through a thought experiment of running NFaaS on existing serverless
platforms to see if the manageability, efficiency, and billing benefits of
serverless hold while meeting the performance requirements of NFs.

\noindent \textbf{Serverless Background.} In FaaS or serverless computing, the user writes a
function, uploads it to the serverless platform and registers for an event
(e.g., object uploads, incoming HTTP request) to trigger function execution.
\emph{The event is the granularity at which the platform does both work assignment (event
routing decisions) as well as billing}. When an event arrives, the platform
routes the event to a compute unit that runs this function. An event may cause
setting up of a compute unit from scratch which involves launching the unit
and downloading the relevant run-time and the function code from a data store;
alternatively, an event may be sent to an already launched and “warmed up” compute
unit. The platform provider can choose from a wide array of available
virtualization technologies to realize the compute unit abstraction, e.g.,
containers, microVMs, or containers-within-VMs. Additionally, the platform
elastically scales computation up/down by based on incoming event rate.

A key aspect of serverless platforms today is that functions are stateless. 
Thus, all state needed to process an event is
read from an external store (e.g., S3~\cite{S3}), and any generated state is
written back to the store. The statelessness simplifies
the elasticity logic as it does not need to worry about state and can
trivially setup/teardown compute units.

Serverless platforms can be viewed to have early roots in Platform-
as-a-service (PaaS) offerings (e.g., Heroku~\cite{heroku}, Google’s App
Engine~\cite{gap} etc) with the key differences being that they support a
broader range of applications, have more fine-grained autoscaling and billing
support, and adopt usage-based
billing instead of time-based (e.g.,~\cite{gap} charges by the hour) , in comparison to PaaS~\cite{berkeley-serverless-report}.


\noindent \textbf{Gaps in serverless platforms today.} While NFaaS atop
serverless platforms, a key design decision that needs to be taken is the {\em event
granularity}, as it decides both the work assignment and billing granularities.
There are two obvious granularities of incoming events - 

\begin{denseitemize}

\item \textbf{Per-packet.} The platform performs work
assignment for every packet, and NF function execution is triggered
for each packet. This mode efficiently utilizes compute units, which helps the
cloud provider, and also ensures that users are billed exactly for just the
compute cycles used. But it comes at the cost of extreme reordering as packets
are sprayed across the compute units in an independent manner. Also,
for stateful NFs, state would need to be accessed via an external store for each packet due to
the stateless nature of functions today, leading to high latencies.

\item \textbf{Per-flow.} The platform 
receives an event when a new flow arrives, which triggers NF function
execution. All packets from this flow are handled by this running function.
This causes no reordering, and state can be maintained/cached locally. But, it
impacts billing, efficiency and performance. It leads to the user being
charged for periods even when no packets are being processed by the function
as it busy-waits for packets pertaining to the still active ``event'' (flow) to arrive.
Also, this mode enforces a long-term commitment between a flow and a running
function, which can lead to overload, affecting performance, or under
utilization, affecting efficiency, depending on the assigned flow's rate.

\end{denseitemize}

In conclusion, naively trying to run NFs atop serverless platforms leads to a
trade-off between the manageability, efficiency, and billing benefits while
meeting our performance targets. These trade-offs are fundamental and
intrinsic to the two gaps present in serverless platforms today, namely: (a) the
tight coupling between workload assignment and billing granularities and (b)
the stateless function abstraction.

\subsection{SNF Key Ideas} 

Given the issues that serverless platforms face today, we now describe our two
key ideas that overcome these limitations and enable us to meet our goals.

\noindent \textbf{1. Decouple work assignment and billing granularities:} As seen above, the
coupling between the two granularities leads to fundamental trade-offs between
our goals. Thus, we advocate for decoupling the two granularities. The key
question that  needs to be addressed is what are the new granularities at
which the platform should operate? We next discuss the granularities that we
choose -  

\noindent \textbf{Billing and programming at per-packet granularity.} Having events as packet arrivals leads to ideal
billing as discussed earlier. Additionally, the notion of having packets as
events also naturally aligns with the way developers typically implement NFs -
take a packet as input and execute the processing function (process\_pkt()).

\noindent \textbf{Work assignment at per-flowlet granularity.} Rather than making decisions at a
flow granularity (as done by NF platforms today~\cite{stratos}), we advocate doing 
it at a {\em flowlet} granularity. A flowlet~\cite{flowlets} is a burst of packets that is separated
from other bursts of packets from the same flow by a sufficient gap called the
flowlet timeout. Acting at this granularity provides more
opportunities to assign/allocate work which limits the negative impacts of
operating at the flow granularity.

While operating at a flow granularity, multiple flows are assigned to the same
compute unit which can lead to under utilization/overload\footnote{Impact
varies depending on whether flow migration is supported or not.} and head of
line blocking (HOL): packets from an earlier elephant flow can cause those from
mice flows later to wait in buffers at the unit, degrading latency (see~\secref{ss:ceval}). On the
other hand, while operating at the flowlet granularity, large flows are
``broken up'' into many flowlets that can now be assigned at many 
units. This mitigates HOL blocking and over-utilization. Also, the
smaller size of flowlets than flows enables better packing of work to compute
units, and hence is better at avoiding under utilization.

\noindent \textbf{2. Eschew complete statelessness and adopt ephemerally stateful functions:} Complete statelessness due to the external store
overheads~\cite{pywren}. While efforts have been made to reduce overheads by
using low latency networking~\cite{statelessNF, chc}, they still experience a
hold up of packets in queues when a flow is reallocated to a new compute unit
(waiting for state to be made available before processing begins). This
exacerbates tail packet processing latencies by 15x (see~\secref{ss:seval}).

At the other extreme is the option of making functions fully stateful by
simply keep state locally at the compute unit and routing all the appropriate
packets to the same unit (similar to~\cite{opennf}). We eschew this option as
it places strict constraints on event routing and defaults to doing flow level
work allocation, which has several negative impacts as discussed above.


Instead, we choose a middle ground. We leverage the fact that all
packets in a flowlet are going to be processed at the same compute unit (as
work assignment is done at flowlet granularity) and maintain state locally
just for the duration of the flowlet’s processing. 
We leverage the flowlet inactivity period to opportunistically  transfer
ephemeral state from one compute unit to another (in case subsequent flowlets of
the same flow are assigned to different compute units) helping us curtail
delays and bring down tail latencies  due state unavailability. With access to such state,
functions now become {\em ephemerally stateful}; this makes packet processing
latencies comparable to when state is maintained locally.


	\section{\name Architecture Overview}
\label{sec:overview}

We now give an overview of \name{}, a serverless NFaaS platform. It has two main components: controller and
NF runtime. The underlying compute units on which the NF logic runs can be
anything - VMs, containers etc.

\noindent \textbf{Controller.} When registered events (packets) arrive at the
platform, the controller forwards them to the appropriate
compute unit where the function (NF logic) executes and the
event is processed. The controller consists of three main modules: (a) the
workload granularization module (WGM), (b) the work assignment module
(WAM) and (c) the state management module (SMM).

When packets from different flows arrive at the controller, the WGM is
granularizes the incoming flows into flowlets.
Specifically, given a packet it determines which flowlet it belongs to (also
involves detecting if this packet starts a new flowlet). If it belongs to an existing flowlet, then it is routed to the appropriate compute unit. If not, the WAM determines the appropriate compute
unit to which this new flowlet should be assigned, so as to 
meet our goals of performance and
efficiency. The SMM encodes a small amount of metadata to each packet -
which compute units use to push/pull state to/from and logical clock to
prevent stale state updates.

\noindent \textbf{NF Runtime.} This runtime realizes the notion of ephemeral stateful
functions (but NF developers are not aware about the ephemeralness) and is
responsible for state management at each compute unit. It transparently handles
updates and transfers among compute units in a peer-peer manner. The state 
management transparently is not visible to the NF developers.

\noindent \textbf{Programming Model.} Given that packet arrivals are events in \name, the NF
developers operate in a familiar model wherein the process\_packet()
routine is called whenever a packet arrives.  Also, the NF runtime
exposes simple put(key, value) and get(key) APIs which the developers use to
access state.

Next we describe the novel compute and state management approaches adopted by \name that maximize utilization without sacrificing performance.
	\section{Compute Management}
\label{sec:compute}

\floatname{algorithm}{Pseudocode}
\begin{algorithm}[t!]

\begin{scriptsize}
\begin{algorithmic}[1]

\State {$\triangleright$ Given a packet P, decide which compute unit to send to}
\Procedure{EventRouting}{Packet P} \label{event_routing}
	\State $\mathbb{M}$ \Comment{Mapping between flowlet and compute unit}
	\State T = \textsc{ExtractTuple}(P)
	\If {\textsc{FlowletDetector}(T, P)} \Comment{Detects new flowlet}
		\State{$\triangleright$ Call \textsc{WorkloadAssigner} as new flowlet is detected}
		\State ComputeID = \textsc{WorkloadAssigner}(T, P) 
	\Else
		\State{$\triangleright$ Use the existing assignment as P is within current flowlet}
		\State ComputeID = $\mathbb{M}$[T] \Comment{}
	\EndIf
	\State \textbf{return} ComputeID
\EndProcedure

\Statex

\State {$\triangleright$ Given a flowlet F and packet P, detect new flowlet}
\Procedure{FlowletDetector}{Flowlet F, Packet P} \label{newflowletdetected}
	\If{$currTime$ - F.\textsc{lastPktTime} $>$ $timeout$}
		\State {$\triangleright$ New flowlet detected as timeout criteria met}
		\State \textbf{return} \textsc{True}
	\ElsIf{P.\textsc{Size} + F.\textsc{Size} $>$ $sizeThreshold$}
		\State {$\triangleright$ New flowlet detected as size criteria met}
		\State \textbf{return} \textsc{True}
	\Else
		\State \textbf{return} \textsc{False}
	\EndIf

\EndProcedure

\Statex

\State {$\triangleright$ Given a flowlet F, assign a compute unit}
\Procedure{WorkloadAssigner}{Flowlet F} \label{}
		\State $\myvec{C}$ \Comment{Candidate compute unit IDs}
		\State $\myvec{G}$ \Comment{Sorted active compute unit IDs}
        \ForAll{$g \in \mathbb{G}$}
		\If{F.\textsc{DemandEstimator(F)} + $g$.\textsc{Load()} $>$ $g$.\textsc{Capacity()}}
        	\State  $score$ = $g$.\textsc{Utilization()} + $\alpha$ * $g$.\textsc{StateExist(F)}
        	\State {$\triangleright$ Add the computed score to $\myvec{C}$}
			\State $\myvec{C}$.\textsc{add(}$g, score$\textsc{)}
		\EndIf
	\EndFor
	\State {$\triangleright$ Pick the compute unit ID which has the maximum score}
	\State \textbf{return} \textbf{max}($\myvec{C}$)
\EndProcedure

\Statex

\State {$\triangleright$ Given a flowlet F, estimate the rate}
\Procedure{DemandEstimator}{Flowlet F} \label{}
	\If{\textsc{IsNewFlow(F)}}
		\State {$\triangleright$ Estimate is average across all flowlets load seen until now} 
		\State \textbf{return} $global average$
	\Else
		\State {$\triangleright$ EWMA(F, ID) = $\delta$ * $F_{ID-1}$.\textsc{Load} + ($1 - \delta$) * EWMA($F_{ID-2}$, ID - 2)}
 		\State \textbf{return} \textsc{EWMA}(F, ID)
 	\EndIf
\EndProcedure

\Statex

\end{algorithmic}
\end{scriptsize}
\caption{\name Compute Management}
\label{alg:compute}
\vspace{-0.2cm}
\end{algorithm}

The \name controller is responsible for the compute management. We
begin by describing the approach adopted by \name to granularize the
incoming workload into flowlets 
(\secref{ss:flowlet}) and then explain the work assignment algorithm 
(\secref{ss:packing}). The overall logic of the WGM and WAM
is shown in Pseudocode~\ref{alg:compute}.

\subsection{Workload Granularization Module (WGM)}
\label{ss:flowlet}

The incoming workload is assumed to be an aggregate of packet flows, which are
identified by the 5-tuples in the packet header. 
Without loss of generality, we assume in this discussion that the
entire packet workload needs to be processed by a single NF type
although we can easily accommodate different NF types.

In \name, when a packet arrives, a new flowlet is detected if one of
the following two criteria is met - (a) if the gap between the current
packet and the previous packet of the same flow is greater than the
flowlet inactivity timeout; or (2) if the size of the existing flowlet
exceeds the flowlet size threshold (lines 14-25 in
Pseudocode~\ref{alg:compute}). We elaborate on the benefits provided
by detecting flowlets in this manner below (\secref{ss:packing}).
Both the timeout and size thresholds are
configurable parameters and poor choices will impact the overall
efficiency and performance. We carry out sensitivity analysis
in \secref{ss:sense-eval}. In case a new flowlet is not detected, 
then the controller forwards the packet to the appropriate compute 
unit (already associated with the flowlet the packet belongs to).

The WGM is also responsible for
estimating the rate of incoming new flowlets, which is needed while making
work assignment decisions. The aggregate workload, and the individual
flows within it, arrive dynamically, and the flows' rates vary. Thus it
is hard to make good resource assignment decisions without having a
reasonable estimate of the incoming flowlet demands. For our
prototype, we estimate the demand of the first flowlet of a flow to be
the average load of all flowlets (across all flows) seen in the
past. For subsequent flowlets, we use an estimate that computes an
exponentially weighted moving average (EWMA) over the previous
flowlet's {\em measured} rate and the previous estimate (lines 40-49
in Pseudocode~\ref{alg:compute}).

\subsection{Work Assignment Module (WAM)}
\label{ss:packing} 

If the WGM has detected the start of a new flowlet, it sends an assignment
request to the workload assignment module (WAM) along with its load estimate.
WAM is responsible for assigning the new flowlet to the appropriate
compute unit running a single NF's
code as in~\cite{aplomb} or a composed NF chain code~\cite{flurries}. In the future, we plan to consider the handling of
NF chains spread across compute units in \name.

Depending on the actual NF logic, the packet processing demands may vary. We
make a practical assumption that each NF compute unit is provisioned with adequate CPU and
memory resources to support a packet workload up to $\mbox{BW}_{max}$bps\footnote{This maximum
rate could also be specified in packets per second at a specific packet size,
say 64-bytes. For the sake of simplicity, and without loss of generality, we
specify this in bps.}. As long as the incoming aggregate rate to the compute unit is less than $\mbox{BW}_{max}$, the NF will be able to provide the requisite performance. Typically, small amounts of overload can be tolerated
when there will be some queuing at the NF and the latency will increase. The
goal of \name is to avoid overload situations even in the face of highly
dynamic workloads.

Our work assignment algorithm greedily packs flowlets 
to active compute units so as to maximize their
utilization. This is analogous to bin packing: balls
are flowlets, and bins reflect the network processing capacity at
compute units (line 34). But the ``greedy'' aspect arises from the
fact that in our approach the compute units are considered in a {\em
  deterministic} sorted order of their IDs; the smallest ID unit with
room is chosen, which leads to units with lower IDs being packed first
(lines 26-39 in Pseudocode~\ref{alg:compute}). This determinism makes
the algorithm simple and easy to implement scalably. Also, we show
that it makes it easy to take state availability into account while
making assignment decisions and accounting for ephemeral state (line
32 in Pseudocode~\ref{alg:compute} - see \secref{sss:algo}); this is
crucial to balance utilization against per-packet latency.

Apart from requiring the load estimate of the flowlet
that is being packed, the algorithm also requires the current load of
the compute units to make decisions (line 31 in
Pseudocode~\ref{alg:compute}). To obtain the current compute load, given that each compute unit is managed
by a single controller, the controller measures the rate at which packets are drained for a
particular compute unit as it is representative of its 
load. 

The controller adds new compute units if the existing ones are saturated, and existing compute units are made inactive if they do
not receive any packets for a fixed amount of time. The start-up times of
compute units also need to be considered while scaling. Given that
there is significant effort ongoing to reduce this overhead~\cite{ed-atc18}, we do not focus on this issue in this paper. Instead, we use the simple strategy of proactively starting them when existing units start to get heavily utilized (say all have load $>90\%$) to mask the overhead.

\textbf{Adversarial Flowlets: } A key issue in packing 
arises when traffic demand spikes suddenly on certain,
or all, flow subspaces. These flowlets that we term as adversarial flowlets, have actual flowlet load that is
significantly higher than the estimate provided by the WGM; in such a
situation the adversarial flowlet can degrade the performance of other
flowlets that are assigned to the same compute unit by building up
queues. If flowlets are detected by just using the inactivity timeout,
the impact of an {\em adversarial flowlet} can last for an arbitrarily long
duration. Thus, in
\name, we bound the negative impact of adversarial flowlets by forking
a new flowlet from the current flowlet if the current flowlet's size
exceeds a predefined size threshold. By bounding size in this manner,
we ensure that adversarial flowlets are drained quickly and their
impact on the overall processing at a compute unit is mitigated. The
new flowlet forked after exceeding the size threshold undergoes the process of
assignment using an updated load estimate, wherein the moving average
(lines 26-47 in Pseudocode~\ref{alg:compute}) accounts for the rate spike
observed in the previous adversarial flowlet.

	\section{State Management}
\label{sec:state}

A stateful  NF's actions on a packet depend on the current state, and for
correct and high performance operation, fast access to correct updated state
information is crucial. NFs may maintain per-flow or cross-flow state. We
focus on per-flow state since it is the common case and plan to consider
cross-flow state in the future. Also, NFs have configuration state which is
often static (e.g., an IDS has string matching rules) and does
not vary at packet-scale timelines. In \name, such state is stored in an
external store and is pulled during the compute unit setup phase leading to no
visible overheads.

In \name, per-flow state management is done {\em transparently using the NF
runtime}.   NF developers build NFs using well defined APIs that are
exposed by the NF runtime, using which they read/update state, and without worrying about state management across compute instances. The run
time makes the current values of state available where needed.

We begin by describing {\em ephemeral state} and how it 
enables fast state operations (\secref{ss:esf}). We then discuss how to
ensure state is available locally at a compute unit even when
subsequent flowlets are assigned to different units. We end
with how updates to stale state information are prevented
(\secref{ss:clocks}).

\subsection{Ephemeral State}
\label{ss:esf}

A compute unit in \name maintains state locally while processing a
flowlet. This state is ephemeral as it is bound to a unit from just
before the first packet of the flowlet is processed till the time the
last packet is done being processed. Once the flowlet has ended, this
state is no longer associated with its compute unit. Ephemeral state
ensures that packets within a flowlet are processed quickly as state
access is always local and fast for each arriving packet.

Ephemeral 
state is initialized when the first packet of a flowlet arrives at a
compute unit as follows: if the flowlet is the first one of the flow,
then the state is set to null; otherwise, if the state has already
been copied over to the compute unit's memory (as described next), then this state value
is used; else, the compute unit pulls state from the remote unit  where
the previous flowlet was processed. The controller sends the
processing location of the previous flowlet of the same flow as
metadata along with the packet belonging to the new flowlet.

\subsection{Peer-to-Peer In-Memory  State Storage}
\label{ss:reactive}

Different flowlets of a flow may be processed by different compute
units depending on the decisions taken by the work assignment
algorithm. This could lead to a scenario where a flowlet $f_1$ of a
flow $F$ arrives at a different compute unit from the one that the
prior flowlet $f_0$ of the same flow $F$ was processed. Clearly state
information is needed at the new compute unit before packet processing
can begin. When state is not available, packets are held up in buffers
at the compute unit until the state is initialized, affecting latency.

With \name, we adopt a \textit{peer-to-peer in-memory state storage
  service} to minimize stalls. Here, ephemeral state is replicated
{\em proactively} in a peer-to-peer fashion by leveraging the {\em
  gaps that exist between flowlets} of a flow. This solution works
well if the amount of per-flow state maintained by an NF is small
enough that it can be transferred during the inter-flowlet gap and not
cause stalls. Luckily, prior work~\cite{statealyzr} has shown that
per-flow state size in commercial NFs like PRADS~\cite{prads} and
Snort~\cite{snort} is under just a few KB for the entirety of a flow's
lifetime; even smaller fraction of this may be updated per
flowlet. Note, however, that proactive replication is unlikely to help with
flowlets that were created from packets exceeding the size threshold (as
opposed to the inactivity timeout).

A key issue is that the above idea requires compute units to
communicate with each other directly. This is a substantial departure from
existing serverless platforms, where units (e.g., lambdas~\cite{AWSlambda}) are disallowed
from communicating with each other, and all communication can happen only via
the external state store.  We do not view this constraint as
fundamental, and for performance reasons, relax it  to enable
communication between cooperating compute units. 

However, to ensure the peer-to-peer in-memory state store is
performant and useful, two key questions need to be addressed - (1)
when should a compute unit proactively initiate state transfer?
and (2) where should it transfer state to?

\subsubsection{When to transfer?} 

Every time there is a period of inactivity in a flowlet, the
compute unit could assume that the flowlet is coming to an end and
initiate state transfer.
However, it is difficult to
accurately predict when a flowlet will  end. Replicating state
whenever there is a small period of inactivity for a flow may lead to unnecessarily doing proactive state
transfers if the flowlet does not end and more packets arrive.
Waiting till the end of the inactivity timeout would default to reactively
pulling the state, which has performance implications. Deciding how early to proactively replicate state has
implications on the additional bandwidth used to transfer
state.

In \name, we proactively replicate state once the period of inactivity
exceeds {\em half} of the flowlet inactivity timeout to balance 
minimizing wait times against making unnecessary state
transfers. In case this flowlet does not end, processing can carry on
without interruption at the primary compute unit which still holds a
copy of the latest state. However, this can lead to inconsistent state
updates, which we discuss and address in~\secref{ss:clocks}. In case a
new flowlet arrives at a new compute unit before the proactive
transfer begins, we first reactively pull relevant state (from the
compute unit with state for the immediate preceding flowlet). If the flowlet arrives while the proactive transfer is occurring, we hold off processing.

\subsubsection{Where to transfer?}
\label{sss:algo}

\floatname{algorithm}{Pseudocode}
\begin{algorithm}[t!]

\begin{scriptsize}
\begin{algorithmic}[1]
\State {$\triangleright$ Given a replication factor R, decide where replication should occur}

\Statex

\Procedure{DeterministicReplicator}{ReplicationFactor R} \label{fixed_rep}
	\State $\myvec{C}$ \Comment{Candidate unit IDs}
	\State $\myvec{G}$ \Comment{Sorted active unit IDs}
	\State {$\triangleright$ Return the first R active compute units}
	\State \textbf{return} $\myvec{G}$[1 : R]
\EndProcedure

\Statex

\State {$\triangleright$ Pick R units using a weighted (inversely to IDs) randomized distribution}
\Procedure{WeightedRandomizedReplicator}{ReplicationFactor R} \label{newflowletdetected}
	\State $\myvec{G}$ \Comment{Sorted active unit IDs}
	\State $\myvec{W}$ \Comment{Weights assigned inversely to IDs}
	\State $\myvec{C}$ \Comment{Candidate unit IDs}
	\While {len($\myvec{C}$) < R}
		\State {$\triangleright$ Pick a compute unit from $\myvec{G}$ where units are weighed by $\myvec{W}$}
		\State replicationSite = \textsc{WeightedRandomizer($\myvec{G}$, $\myvec{W}$)}
		\State $\myvec{C}$.\textsc{add(}replicationSite\textsc{)}
	\EndWhile

	\State \textbf{return} $\myvec{C}$[1 : R]
\EndProcedure

\Statex

\end{algorithmic}
\end{scriptsize}
\caption{\name State Management}
\label{alg:state}
\vspace{-0.3cm}
\end{algorithm}  

Once it is time to proactively push state, the runtime at a
compute unit needs to decide where to replicate state.   A
strawman solution would be to broadcast to all other active 
units but this has the overhead of doing unnecessary
transfers. Instead, in \name, the controller estimates the top K
compute units where the next flowlet of this flow could likely be
assigned to and it keeps track of this information for each
flowlet. The reason for picking the top K and not the exact one is
because it is not possible to know ahead of time as to where the next
flowlet would be assigned. The reason is that a compute unit that is
available currently may be saturated by the time the new flowlet
arrives (due to flowlets of other flows being assigned in the
interim). The question is how to pick the ``top K'' such that the probability of the compute unit chosen by the WAM for the next flowlet already having the necessary state is high.

We could replicate state to the K  least loaded units, 
expecting that the WAM would assign the next flowlet to
them. However, the load can change by the time the next
flowlet of this flow starts. Also, implementing a load-aware strategy
is complex, as we need up to date load information at scale.

Since the WAM deterministically processes compute units (lines 2-7 in
Pseudocode~\ref{alg:state}), one simple load-unaware strategy is to
pick the least K ID compute units, i.e., units with IDs from 1 to K, to replicate to (the WAM would
preferentially allocate a new flowlet amongst these). But doing this
for every flowlet's replication would render proactive replication
ineffective when the least K ID units become overloaded, which is
likely especially for a small K. In such cases, future flowlets are
assigned outside these K units, and thus they would have to pull state
reactively.

\name uses a simple variant of the above strategy that allows for some error in
the estimated location where a future flowlet goes to. We pick 
the top-K compute units to replicate state to,  with probability inversely proportional to
the units' IDs (lines 8-19 in Pseudocode~\ref{alg:state}). Doing so ensures we
pick the lower ID units' with higher probability as is done by
WAM.

The next question is how should the controller make the next
assignment decision to account for state availability and maximize the
potential benefits of proactive replication? A strawman solution would
be for the controller to check if any of the K compute units (which
have the required state) could handle this flowlet.  If yes, the
flowlet is assigned to one of the units in question and processing can
proceed without any wait time. If not, then we assign the flowlet to
an available compute unit and the state is pulled
reactively. Unfortunately, this approach ignores load, which affects
utilization.  It can cause compute units to become fragmented with many compute units poorly utilized.

Instead we extend the work assignment algorithm to make decisions
using a weighted scoring metric (line 32 in
Pseudocode~\ref{alg:compute}) for choosing from the available compute
units using both utilization and state availability. The weighted
metric is ${S = utilization + \alpha \times \beta}$, where $\beta$ is
1 if the compute unit has the replicated state, otherwise it is 0.
$\alpha$ is a balancing knob between 0 and 1, and balances utilization against
proactive benefits: $\alpha = 0$ results in the controller
making assignment decisions to improve utilization (and ignoring
state) and $\alpha = 1$ biases more in favor units where replicated state
is available.

\subsection{Preventing Updates on Stale State}
\label{ss:clocks}

While the above techniques minimize packet wait time, we need to
ensure that a flowlet does not make updates on stale state that is
present at its corresponding compute unit. This can occur when the
optimistic approach of using half the flowlet timeout as the deadline to proactively
replicate state from an old to a new unit was erroneous in
assuming a flowlet would end. Here, the NF runtime would proactively
copy state, but a few lingering packets from the original flowlet
continue to arrive at the old unit and update state there. State
updates due to such packets should be reflected in the state copied
over to the new location before {\em any} processing begins there.

To prevent a new flowlet from acting on stale per-flow state at the new
unit, we introduce the notion of {\em monotonically increasing logical
clocks} for each packet of a flow. These are assigned by the 
controller. Each packet carries its logical clock as metadata. This
prevents flowlets from making update on stale state in the following manner.
The NF runtime tags the state that is proactively replicated with the logical
clock of the last packet of this flow that was received by
the old unit. When a new flowlet of this flow arrives at the new compute
unit, before making updates to the state, the NF runtime verifies if the
latest state is available by checking the logical clock of the packet (i.e., first packet of the new flowlet) is one more than
the value with which the copied-over state is tagged; if not, state update due
to the new packet is stalled, fresh state pulled reactively, and then the
update  proceeds.

The above technique also works in the rare event of packets arriving out of
order. As is done today, if the NF logic requires packets to be processed
in order, then the NF developer needs provide appropriate reordering logic.
This typically involves storing the out-of-order packet until the intermediate packet
arrives and then processing them in order. Thus, out-of-order packets become a
part of the ephemeral NF state (which is tagged appropriately to prevent stale
updates as described above) and are processed per the logic defined. 

	\section{Fault Tolerance}

Given that NFs are stateful and the latest state is required for
correctness, we need to ensure that \name is fault tolerant in maintaining per-flow state.
When the originally assigned (or primary)  compute unit for an NF fails while
processing a flowlet, a new recovery unit takes over the flowlet's
remaining processing. The key property we desire is that the per-flow
state initialized at the recovered unit have the same value as
under no failure. We assume the standard fail-stop model in which a
compute unit can crash at any point and that the other units
in the system can immediately detect the failure.

Traditional recovery mechanisms do not work in the NF context due to the
performance constraints as well as the presence of non-deterministic state
update operations in the logic (e.g., ``random'')~\cite{ftmb}. Thus, we sketch a solution that builds on prior work on
fault tolerance for NFs~\cite{chc,ftmb}. While past solutions covered
general NF chains, where different types of NFs were deployed across
different units, our solution is simpler owing to our design choices.
Specifically, recall that \name handles
only composed chains that are run in the same unit. Also,
our NF units have a single packet processing thread. This leads to
the following approach.

In \name each NF compute unit is coupled with a separate output logger 
(OL) unit, which is launched on a different physical machine.  
Once a packet has been processed by an NF, 
the packet along with its {\em state delta} is
sent to its OL; delta is the change to the value of a piece of
state. It is the responsibility of the OL to use the delta to locally
update state it maintains in memory for the NF, and only then forward
the packet externally. Thus, the OL maintains a consistent copy of the
state of the primary NF compute unit. Note that the output logger can
also be implemented using the same packet-based programming model as
NFs in \name. 

We assume that the NF and its OL do
not fail at the same time. To protect against simultaneous failures, the state must
be replicated at multiple OLs before the packet is released. This increases overhead and we do not consider it here.
If warranted by specific use cases, additional OL resources can be added to ensure correctness even under simultaneous failures.

A recovery unit can take over a failed unit's processing by pulling
the state from the associated OL. On the other hand, if the OL fails,
then another OL is brought up and initialized by pulling state from
its associated NF compute unit. The controller provides the necessary
metadata to the failover NF/OL to pull state from the relevant compute
unit\footnote{We differ from~\cite{chc,ftmb} in that
  we don't need an input logger. They had
  multi-threaded NFs, and complex cross flow state, recovery of which
  requires packet replay from an input logger. Recovery of per-flow
  state simply requires state copy from the OL.}.

While this approach would generally provide the property we desire, a corner
case may still occur: say a packet has been processed by an NF (the
state has been updated at the NF), but the packet along with the state
delta gets lost en route to the OL, as well as the NF unit
fails. During recovery, the failover NF unit would be initialized
based on the state pulled from the OL (which does not have the latest
state of the NF prior to the failure, because it does not reflect the
update made by the lost packet). Nevertheless, using this state offers
reasonable semantics: it is no different than the state value when the
lost packet is dropped by the network {\em before} it is processed by
the original NF unit. The lost packet does not reach the
destination and the sender will eventually retransmit it.

If an OL fails before transmitting a
processed packet, is not different from the packet being dropped by the
network between the OL and the destination. The NF unit
still has the most up to date state; a recovered OL can pull this
state. 

The above protocol provides state fault tolerance but at the
cost of an additional latency of half an RTT, and the additional
bandwidth usage to ship state deltas to OLs. We believe that this overhead 
is reasonable, especially given that
 prior work~\cite{statealyzr} has shown that the deltas across
 successive updates to state objects are typically small.

Finally, to handle controller failure, given that it is
stateful (e.g., logical clock, flowlet to compute unit
mapping), we can write the state to an external store (as 
in~\cite{chc}) and have the new controller read the latest values from the
external store. Alternatively, the operator may opt to replicate the
controller using Apache ZooKeeper~\cite{zookeeper}.
	\section{Controller Scalability} 
\label{sec:scale}

Though the latency overhead introduced by the controller is
minimal~\secref{ss:overheads}, as the input workload scales (e.g., 100
Gbps), having a single (even powerful) controller would lead to it
being a bottleneck eventually. Thus, to support large scale workloads,
we would need to have multiple controllers. One approach is for
controllers to operate on dedicate sets of compute units. But this
impacts compute efficiency due to resource fragmentation as
controllers do not share units. Alternatively, the
controllers can share the underlying compute units by using a state
store to share compute unit load information.  But imposes the
overhead of coordinating over store access for every work allocation
decision, i.e., {\em every flowlet}.

Instead, in \name, we use a global resource manager (RM) that manages
a pool of compute units.  The various controllers ask for the required
processing capacity based on the load seen in the last epoch (say
100ms). The RM allocates the requested capacity, which can be
fractional; e.g., the RM could allocate 2.5 units, which requires
spinning up 3 units with the full first two units and half the
capacity of the third unit allocated to the requestor.
When load in the current epoch is nearing requested capacity, the
controller requests for more capacity so as to
avoid performance degradation.
The controllers give back
resources once they become inactive as described earlier. The RM ensures that resource fragmentation across controllers
is reduced as it strives to pack (fractional) units to their capacity.

	\section{Implementation} 
\label{sec:impl}

We built our prototype from scratch in C++ (20K LOC) rather than building off
existing platforms such as AWS Lambda due to their blackbox
nature~\cite{hotnets-18, wang-atc18}. It consists of -

\noindent\textbf{RM.} Implemented as a standalone process, it establishes TCP connections with controllers and handles resource requests.

\noindent\textbf{Controller.} Implemented as a multithreaded process, it
establishes TCP connections with compute units and runs the compute
management algorithm. It measures the compute unit load
by monitoring the rate at which packets are drained. It measures this load
at fixed buckets (of 500us) and considers the
load over multiple buckets when packing (last 200 buckets, 
i.e., the last 100ms). The $\#$buckets considered indicates the minimum
time for which a change in traffic pattern should exist for the system to
react. We choose the above values because smaller values
made our system unstable by reacting to minor bursts, and larger values cause
it to react too slowly.

\noindent\textbf{External Datastore and NF Runtime.} Implemented as 
multithreaded processes. The former is used to hold NF configuration state and the
latter does ephemeral state management.
Packet reception, transmission, processing and datastore connection are
handled by different threads. Protobuf-c~\cite{protobuf} is used to encode and
decode state transferred between units. Also, the NF runtime exposes APIs using which we reimplemented five different NFs of varying
complexity -

\noindent\textbf{NAT.} Performs address translation and the list of available
ports is the NF configuration state. When a new connection arrives, it obtains
an available port and it then maintains the per-connection port mapping.

\noindent\textbf{LB.} Performs hash-based load balancing. The servers' list
constitute the NF configuration state. When a new connection arrives, it
obtains the server based on hash, and then maintains per-connection (a)
server mapping and (b) packet count.

\noindent\textbf{IDS.} Monitors packets using the Aho-Corasick algorithm~\cite{aho-corasick} for signature
matching. The string matching rules (e.g., Snort~\cite{snort} rules) constitute the NF configuration state.
Also, the NF maintains per-connection automaton state mapping.

\noindent\textbf{UDP Whitelister.} Prevents UDP-based DDoS
attacks~\cite{bohatei} by recording clients who send a UDP request.

\noindent\textbf{QoS Traffic Policer.} Implements the token bucket algorithm. The per-connection (a) committed rate and (b) token
bucket size constitute the NF configuration state. Also, the NF maintains per-connection mapping of (a) time since previous packet and (b) current available tokens.
			
	\section{Evaluation}
\label{sec:eval}
We evaluate \name to answer the following questions:

\begin{denseitemize}

\item Can SNF provision compute as per the incoming traffic demand at fine time scales? Do we meet our goal of maximizing utilization without sacrificing performance?
\item Does proactive state replication help curtail tail latencies? 
\item How does \name perform when adversarial flowlets occur?
\item How quickly can \name recover in the presence of failures? 
\item Is \name able to reduce resource fragmentation when multiple controllers are being used? 
\item How does \name perform with different system parameters? 

\end{denseitemize}

\noindent \textbf{Experimental Setup:} We use 30 CloudLab~\cite{cloudlab} servers each with
20-core Intel Skylake CPUs and a dual-port 10G NIC. The \name RM and controller run
on dedicated machines. The controller receives the replayed traffic from traces (details below) while the compute
units run within LXC containers~\cite{lxc} on the remaining
machines. For all our experiments, we use one controller and each compute unit is configured to process
incoming packets at $\mbox{BW}_{max}$=1~Gbps, enabling 10 compute
instances per machine.  The default parameters are: flowlet inactivity timeout
$T = 500\mu$s, the flowlet size threshold $B = 15KB$, the balancing knob
$\alpha = 0.25$ and the replication factor $K = 3$.

\noindent \textbf{Real Packet Traces:} We use two previously collected packet traces on the WAN
link between our institution and AWS EC2 for a trace-driven evaluation of our
prototype. One trace has 3.8M packets with 1.7K connections whereas the other
trace has 6.4M packets with 199K connections. The median packet sizes are
368~Bytes and 1434~Bytes. All the experiments were conducted on both the
traces with similar results; we only show results from the latter trace for
brevity. Given that the load of the collected traces was not high,
we scale the trace files by reducing the packet inter-arrival times. 

\subsection{Compute Management Performance}
\label{ss:ceval}

 \begin{figure*}[t]
 	 \captionsetup[subfloat]{captionskip=-5pt}
 	\centering 
 	\subfloat[][Vanilla Flow]{%
 		\includegraphics[width=0.25\textwidth]{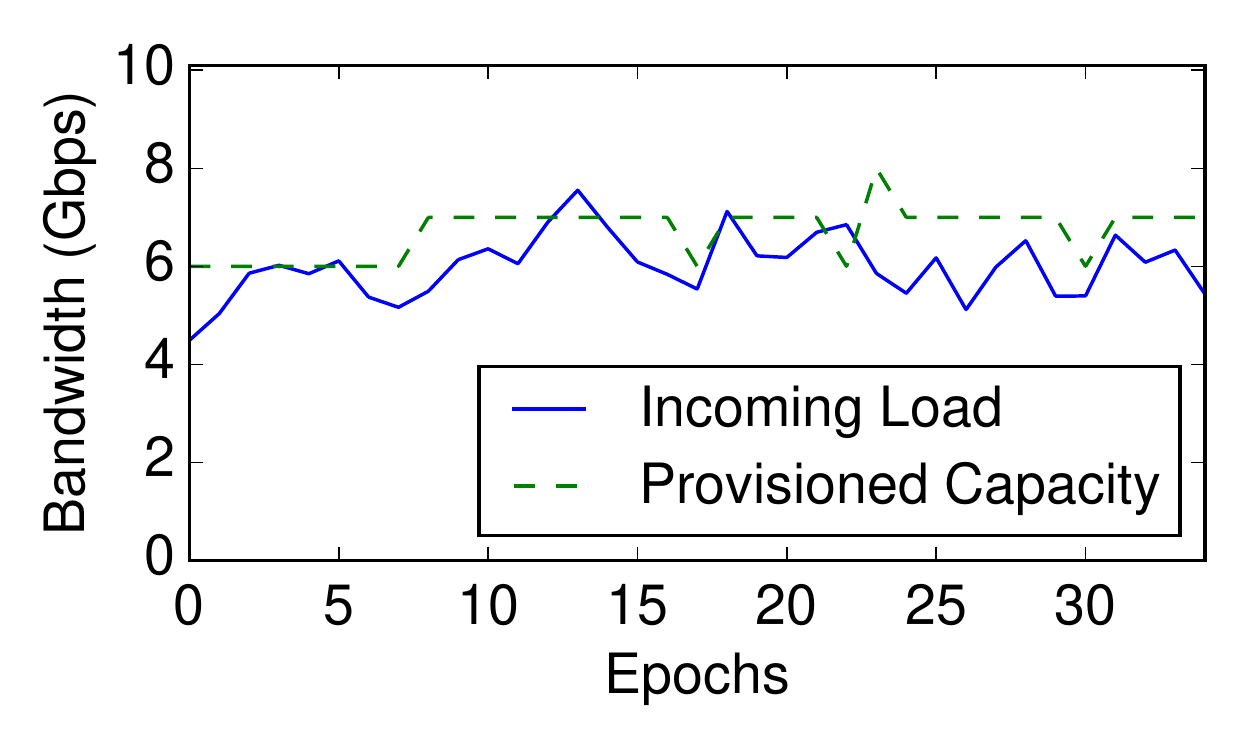}
 		\label{compute-flow}
 	}
 	\subfloat[][Smart Flow (100ms)]{%
 		\includegraphics[width=0.25\textwidth]{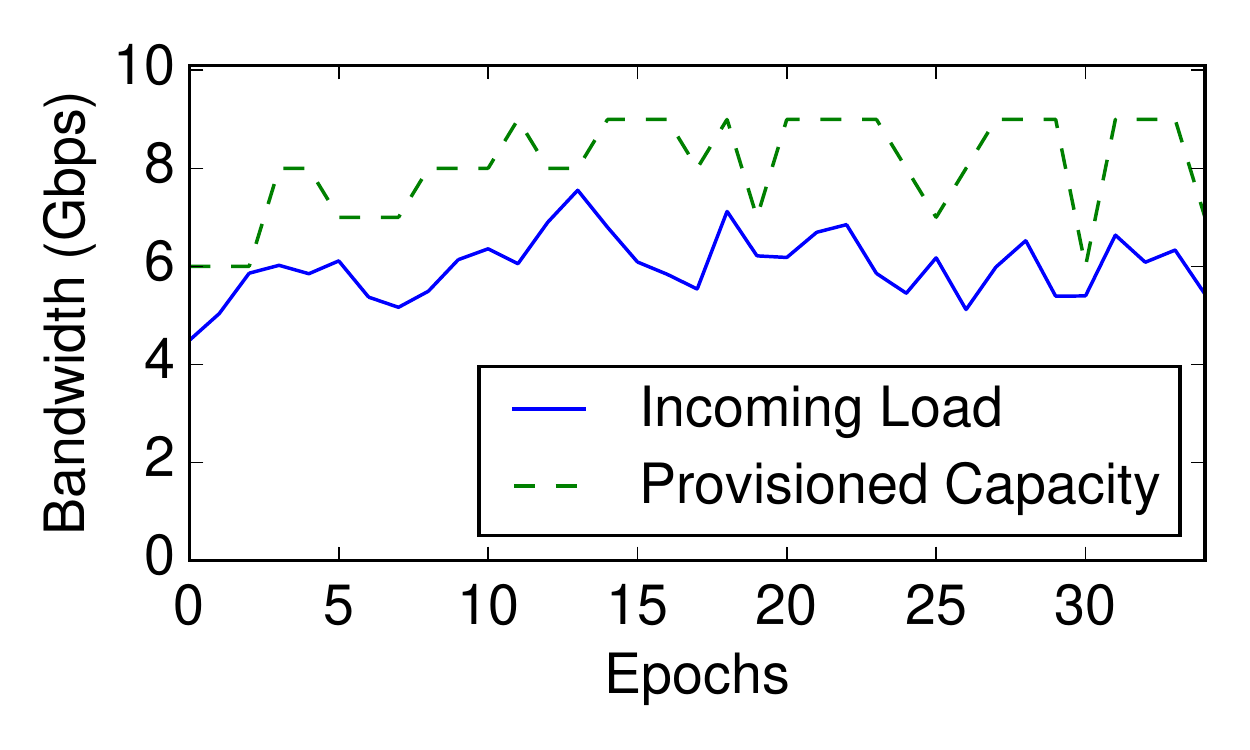}%
 		\label{compute-smart-flow-100}
 	}
 	\subfloat[][Smart Flow (50ms)]{%
 		\includegraphics[width=0.25\textwidth]{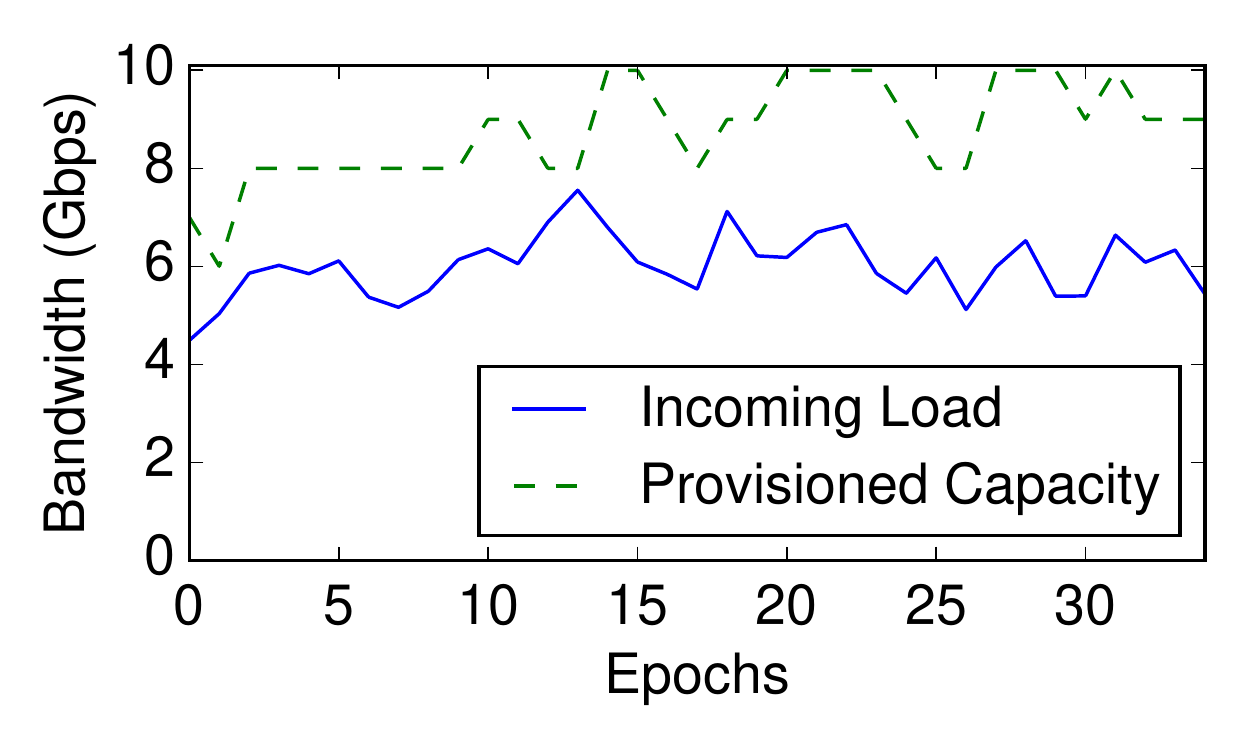}%
 		\label{compute-smart-flow-50}
 	}
 	\subfloat[][Flowlet]{%
 		\includegraphics[width=0.25\textwidth]{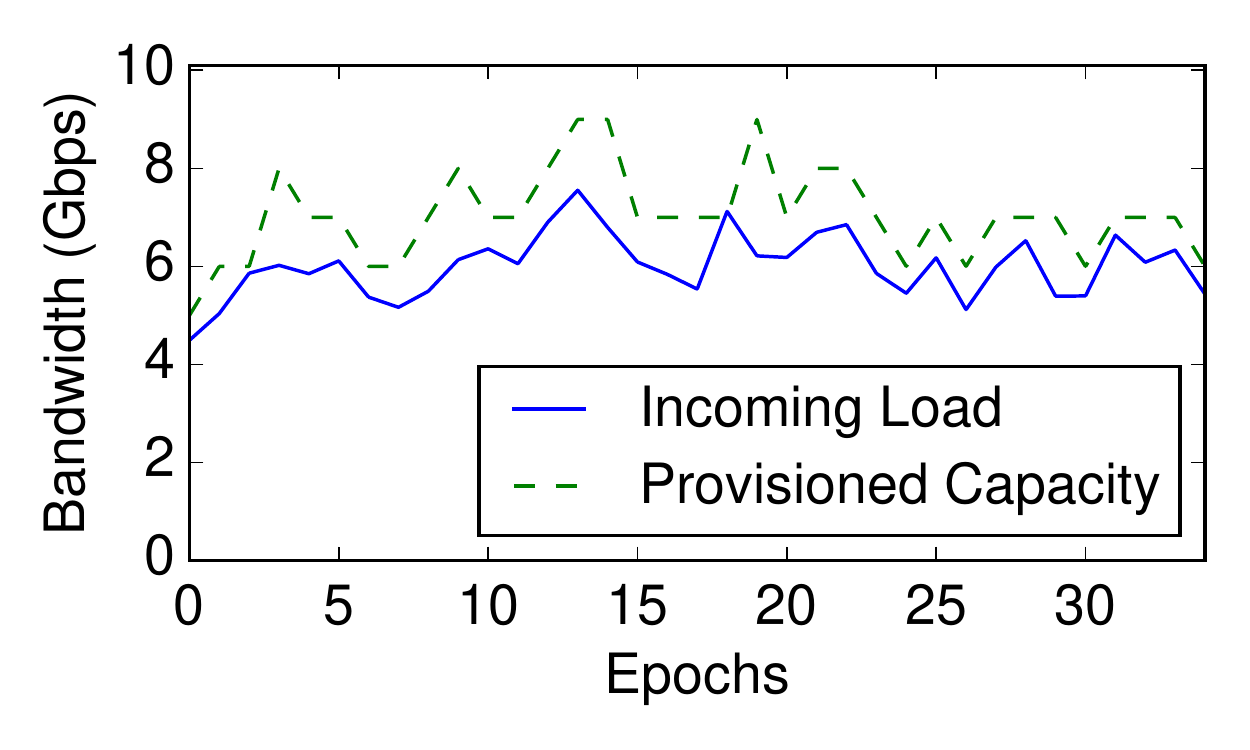}%
 		\label{compute-flowlet}
 	} 	
  	\vspace*{-2mm}
 	\caption{ \footnotesize Compute provisioning across various work allocation modes}
 	\vspace*{-7mm}
 \end{figure*}

  \begin{figure*}[t]
  	 	 \captionsetup[subfloat]{captionskip=-5pt}
 	\centering 
 	\subfloat[][Vanilla Flow]{%
 		\includegraphics[width=0.25\textwidth]{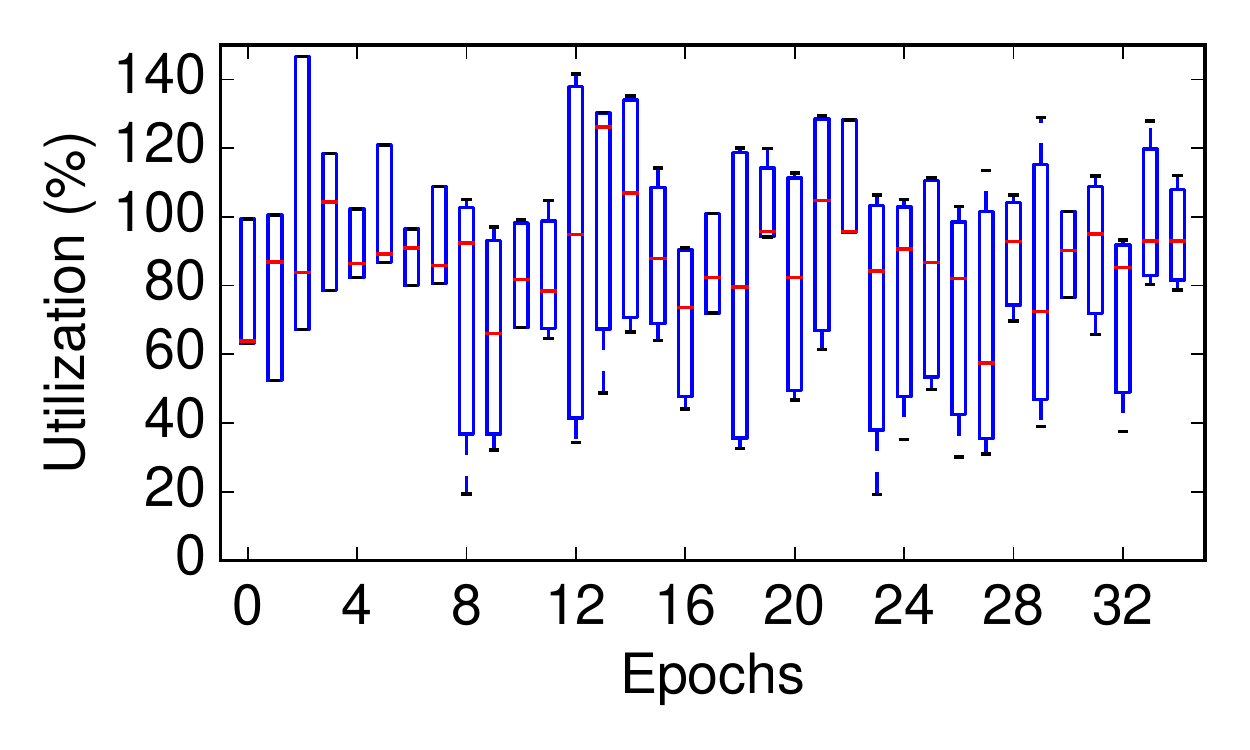}
 		\label{util-compute-flow}
 	}
 	\subfloat[][Smart Flow (100ms)]{%
 		\includegraphics[width=0.25\textwidth]{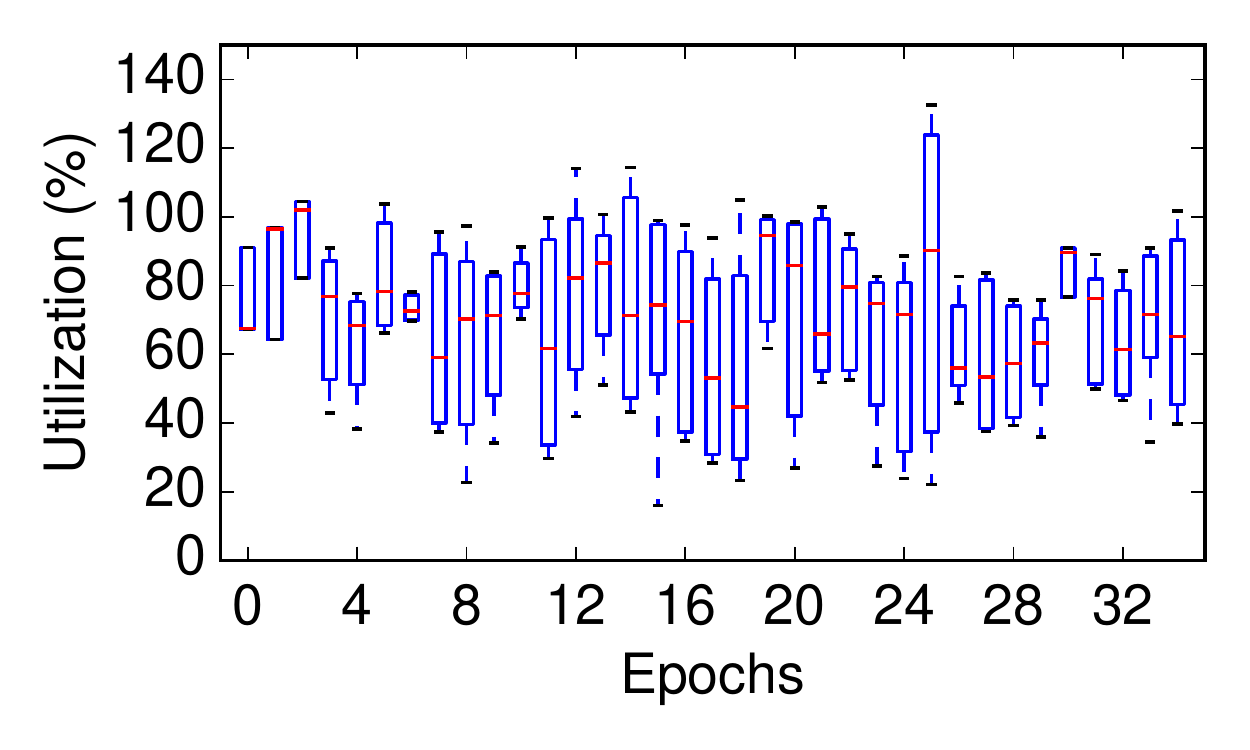}%
 		\label{util-compute-smart-flow-100}
 	}
 	\subfloat[][Smart Flow (50ms)]{%
 		\includegraphics[width=0.25\textwidth]{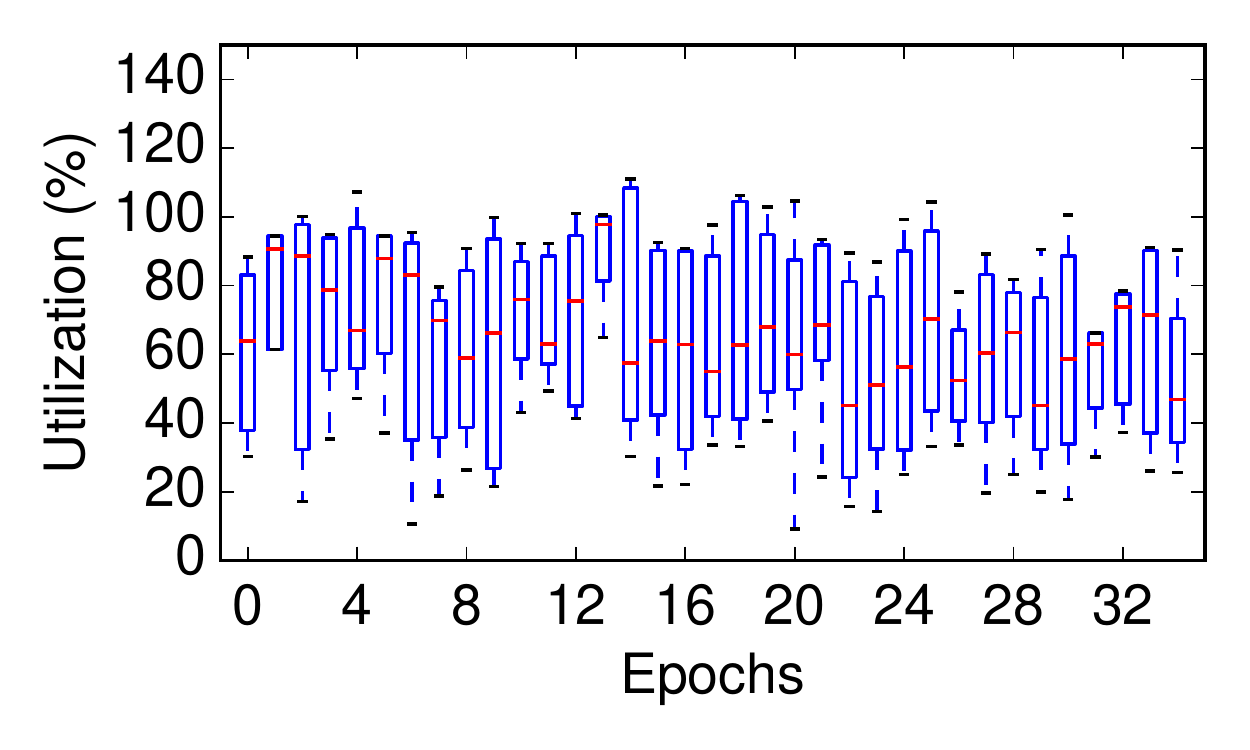}%
 		\label{util-compute-smart-flow-50}
 	}
 	\subfloat[][Flowlet]{%
 		\includegraphics[width=0.25\textwidth]{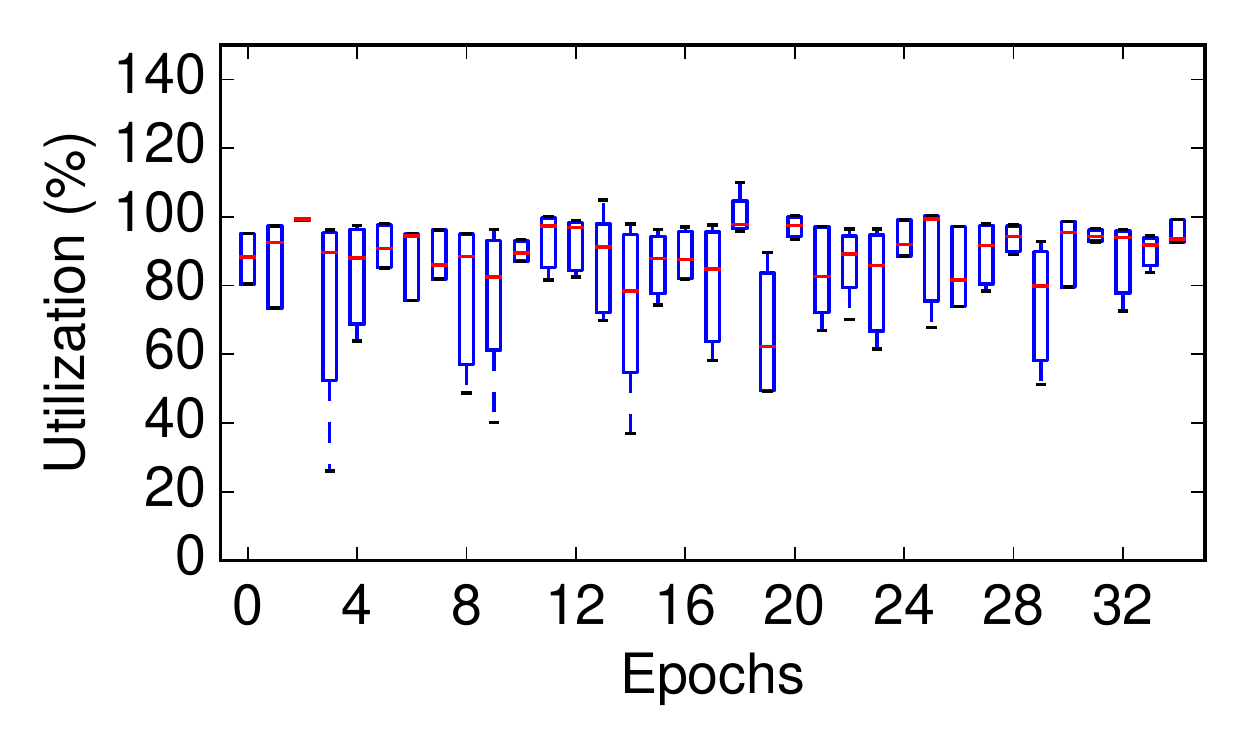}%
 		\label{util-compute-flowlet}
 	} 	\\
  	\vspace*{-2mm}
 	\caption{ \footnotesize Epoch-wise utilization distribution of the active compute units across various work allocation modes}
 	\vspace*{-5mm}
 \end{figure*}

We first evaluate \name's
approach of performing flowlet-level work allocation. We measure (1) the
provisioning efficiency with changes in traffic demands by
recording the number of active compute units at 100~ms time intervals
(referred to as an epoch hereafter), (2) the NF packet processing latencies
incurred and (3) compute unit utilizations. 

We compare against
two other baselines that act at the flow-level: (1) \textbf{Vanilla Flow Allocation:} work
allocation is done when flows arrive, and once a flow is assigned to a compute
unit, it is associated with that unit for its entire lifetime. This mimics
existing out-of-the-box work allocation techniques (when optimized state reallocation schemes~\cite{opennf} are not used).
(2) \textbf{Smart Flow Allocation (X ms):} work allocation is done when flows arrive, and if required, flows
are reallocated every X ms to avoid overload/underutilization at any compute unit. This is
similar to a work allocation scheme that uses state reallocation schemes and
thus supports flow migration~\cite{opennf}.

Figs.~\ref{compute-flow}-\ref{compute-flowlet} show a runtime
snapshot of \name's provisioned bandwidth and the packet processing demand\footnote{Since each compute instance has $\mbox{BW}_{max}$=1~Gbps, the
provisioned bandwidth is 1x \#instances.}. We see that acting on
flowlets enables \name to closely match the incoming load,
which is not the case when acting at the flow granularity, irrespective of
whether flow migration is supported or not.

In the vanilla allocation mode, we see that the system does not
adapt well to the incoming load as it can react only when new flows arrive.
In the smart flow allocation mode, when we reallocate, if
required, every X ms (X being 50ms or 100ms) the system is more adaptive in
comparison to the vanilla flow mode as it gets more opportunities to
reallocate flow. Acting at the flowlet
granularity gives us $\textbf{3.36X}$ more opportunities to assign work as
compared to the alternatives, enabling \name to better react to variations in
the incoming load.

Additionally, when operating in the smart flow mode, there is
over provisioning of units (greater when we are more aggressive to reallocate,
i.e., smart flow (50ms)) due to the poor packability of flows which are larger
work allocation units (in comparison to flowlets). 
However, even when acting in the flowlet mode, at times, additional 1-2
compute units are used in order to extract the benefits of proactive
replication (\secref{ss:seval}).
\begin{figure*}
	\begin{minipage}{0.32\textwidth}
		\centering
			\includegraphics[width=1.0\textwidth]{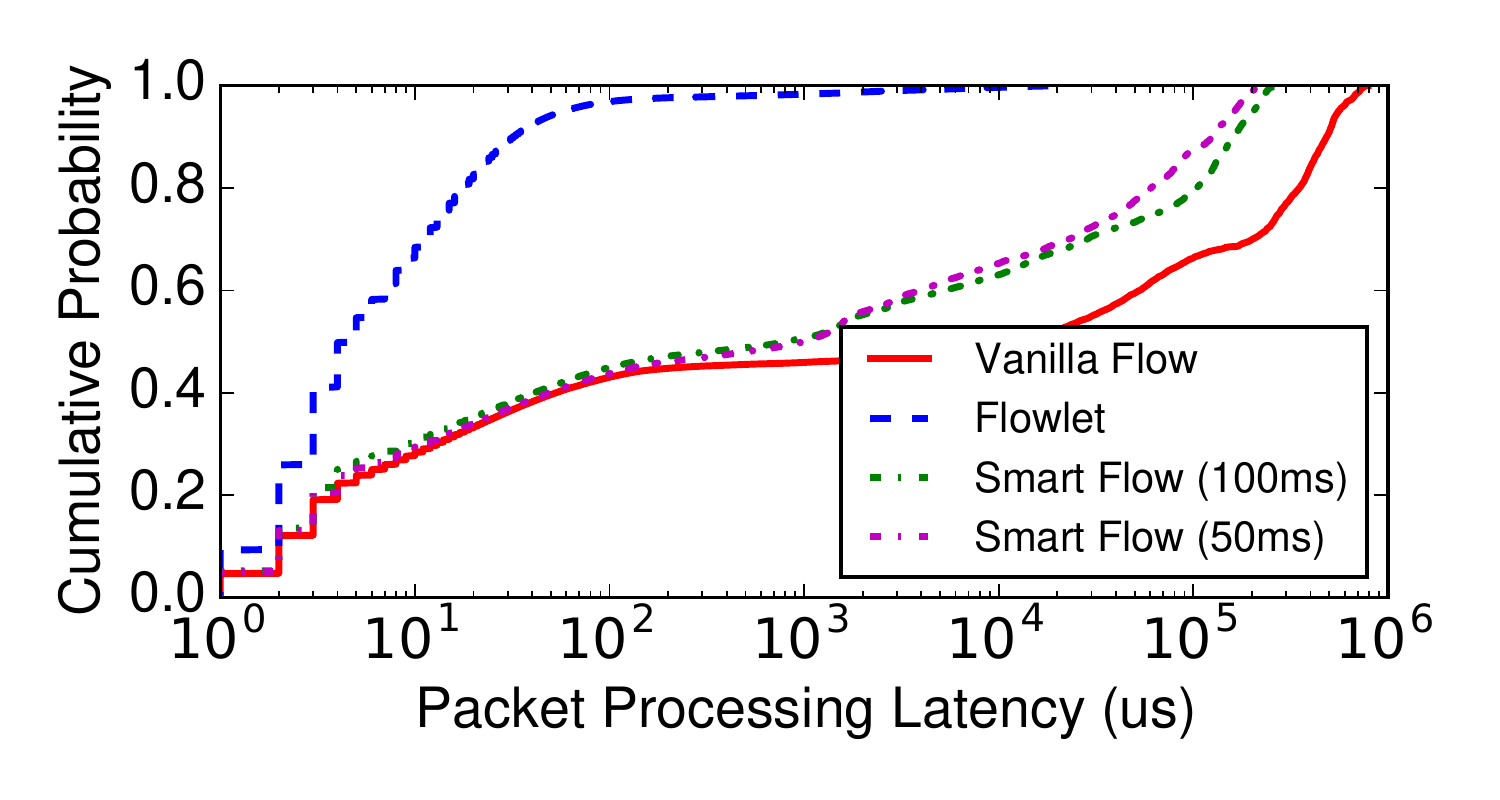}
 	\vspace*{-7mm}
		\compactcaption{ \footnotesize Pkt. proc. latencies (NAT) across various work allocation modes}
		\label{pkt-processing-with-queueing}
	\end{minipage}\hfill
	\begin{minipage}{0.32\textwidth}
		\centering
			\includegraphics[width=1.0\textwidth]{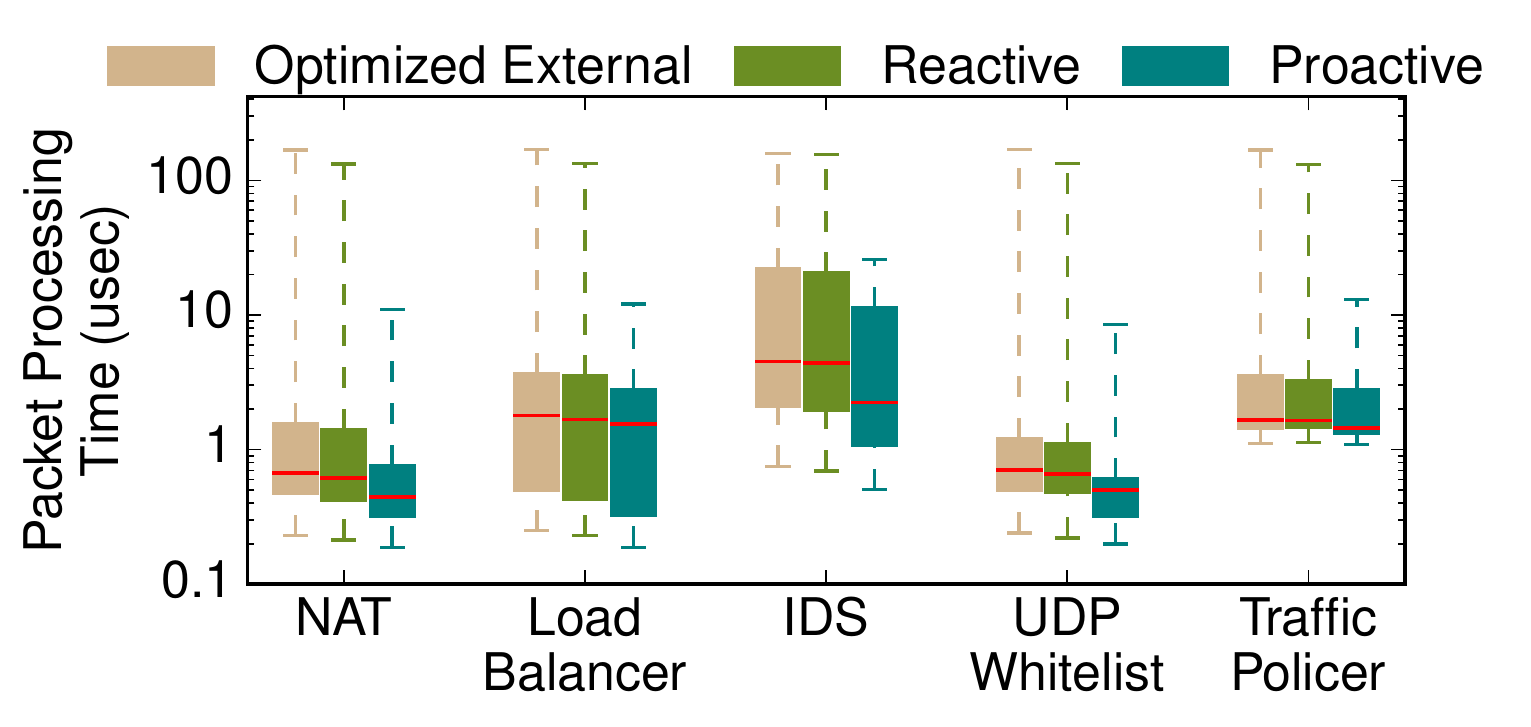}
			 	\vspace*{-5mm}
		\compactcaption{ \footnotesize Pkt. proc. Latencies (1-25-50-75-99\%-iles) for different storage modes}
		\label{pktprocessing-without-queueing}
	\end{minipage}\hfill
	\begin{minipage}{0.32\textwidth}
		\centering
			\includegraphics[width=1.0\textwidth]{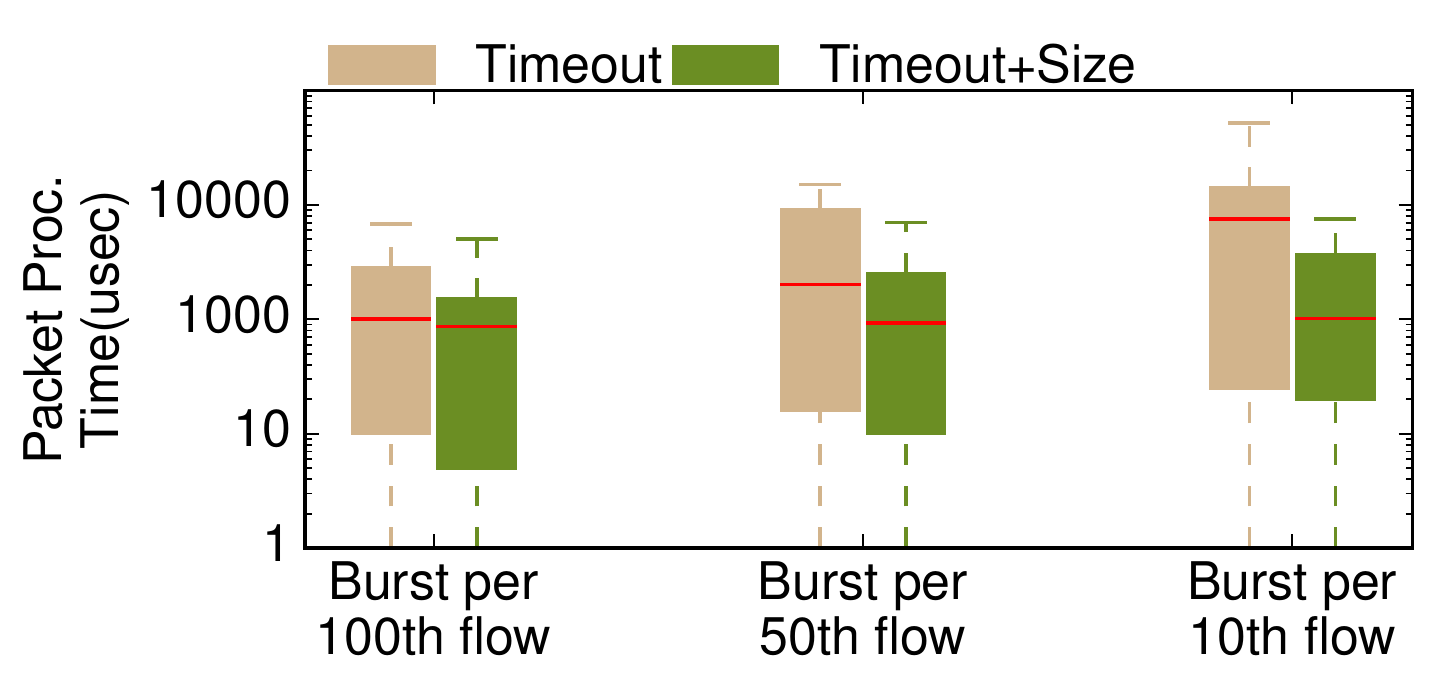}
			 	\vspace*{-5mm}
		\compactcaption{ \footnotesize Pkt. proc. Latencies (1-25-50-75-99\%-iles) in the presence of adversarial flowlets}
		\label{high-load}
	\end{minipage}
	 	\vspace*{-1mm}
\end{figure*}

\noindent \textbf{Packet Processing Latencies.} Fig.~\ref{pkt-processing-with-queueing} shows
that the packet processing latency for the NAT NF while using the vanilla flow
mode is significantly worse in comparison to the flowlet mode: the 75th\%-ile
latency is 275.4ms, which is \textbf{19.6K} times worse than for flowlet mode.
Once flows are pinned to a compute unit, the association continues until flows
end and the presence of elephant flows (5\% of the flows in our trace have a
size greater than 10KB) causes input queues at compute units to build up. The trends for the other NFs are similar.

In the smart flow mode when we reallocate every 100ms (50ms), the latency is
still worse than flowlets: the 75th\%-ile latency is 64.5ms (41.1ms), which is
\textbf{4.6K (2.9K)} times worse than the flowlet mode. This is due to (1) the
mode being unable to handle overloads that occur at lower time-scales than the
reallocation frequency (50ms or 100ms) and (1) hold up of packets once
reallocated until the relevant state is pulled from the prior compute unit. In
the flowlet case the 99\%-ile latency is 2.8ms, while the median is 5$\mu$s.
The tail is contributed by micro bursts\footnote{Since the prototype cannot detect traffic changes that last $<$ 100ms} leading to queuing occurring at the
compute units as well as due to flowlets for which the NF runtime has to
reactively pull state from the previous compute units where the prior flowlet
in the flow was processed. 

\noindent \textbf{Utilization.} While we have seen that operating in the flowlet mode has the
best performance, we need to verify that this improved performance is not
coming at the cost of simply using more compute units. To do so, we delve deeper and
look at the epoch-wise distribution of the active compute unit utilizations
under the various modes (see Figs.~\ref{util-compute-flow}-\ref{util-compute-flowlet}).

As expected, while operating in the vanilla flow mode we experience maximum
number of overloaded compute units as the system is the least reactive.
Interestingly, in the smart flow (100ms) mode, even though over-provisioning
occurs, we do see certain compute units being overloaded and this is due to
the fact that the system can react only every 100ms. Consequently, in the
smart flow (50ms) mode, we see lesser overload. Moreover, in all these modes
wherein we act the flow level, we do experience more underutilization as well
due to the poor packability of flows. 

On the other hand, operating at the flowlet granularity rarely experiences
overload as we get far more opportunities to react and has lesser
underutilization as flowlets being smaller work allocation units pack better
in comparison to flows.

\subsection{State Management Performance}
\label{ss:seval}

We now evaluate  \name's approach of proactively
replicating ephemeral state. We
compare it against two other baselines: (1) Optimized External: state is
proactively pushed (rather than waiting for the flowlet end) to an external
in-memory store and is read at the beginning at the flowlet. This baseline is
an optimization to how state is transferred across compute units in today’s
serverless platforms\footnote{Today serverless platform don't write to
external stores by default; such stores would have to be provisioned by the
application.}. (2) Reactive: state is pulled on the arrival of a flowlet from
the previous compute unit that the flow was processed at. 
 
We measure the per-packet
processing latencies for the various NFs (Figure~\ref {pktprocessing-without-queueing}). 
For the NAT, the median latencies across the three modes, external, reactive and proactive are more
or less similar (0.67$\mu$s, 0.61$\mu$s and 0.44$\mu$s) due to the fact that
state for most flowlets is eventually locally available in all the three
models. However, the tail latencies improve while shifting from the external
to reactive and finally to proactive (168.18$\mu$s,
132.74$\mu$s and 11.01$\mu$s respectively). In both the baselines, 
upon arrival of the new flowlet, the updated state needs to be pulled from the
external store and the previous compute unit respectively, thus the latencies
are dominated by the network RTT~\footnote{The tail latencies in case of using
an external store are slightly higher than when reactively pulling state in a
peer-peer fashion as there may be scenarios where in a flowlet makes a
reactive request to the external store and its not available, and thus has to
wait longer}. In the proactive mode, state is made available prior to the
arrival of a new flowlet (unless there is a delay due to network anomalies or
the flowlet has been scheduled to a unit which does not have replicated state)
due to which processing is not stalled due to state unavailability. Similar
latency trends are noticed for the other NFs.

\noindent \textbf{Proactive State Replication:} We carry out deeper analyses to understand 
where the benefits of proactive replication arise from. For the flowlets that were
assigned to different units than the immediate previous units
for the various NFs, proactive state replication ensured that 90.43\% of flowlets
(on average across the NFs) were able to proceed seamlessly without any wait
time whereas the remaining 9.57\% reactively pulled state. This indicates that
proactive replication comes into effect the majority of time helping to vastly
reduce the tail.

Thus, with state optimizations in place, \name can achieve median latencies
similar to when state is maintained locally which reduces the tail in comparison
to existing alternatives. 

 \begin{figure*}[h]
 	\centering 
 	\subfloat{%
 		\includegraphics[width=0.28\textwidth]{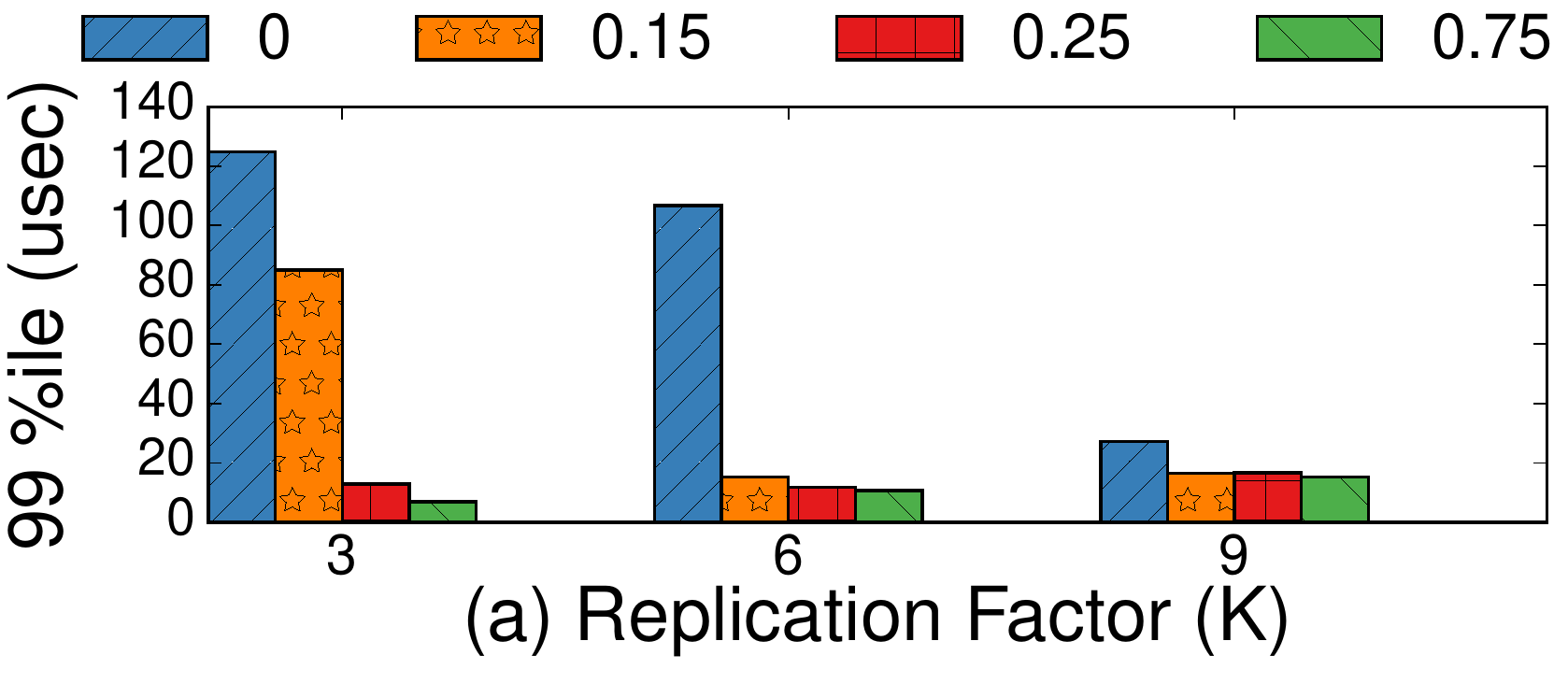}
 		\label{sense-l}
 	}
 	\subfloat{%
 		\includegraphics[width=0.28\textwidth]{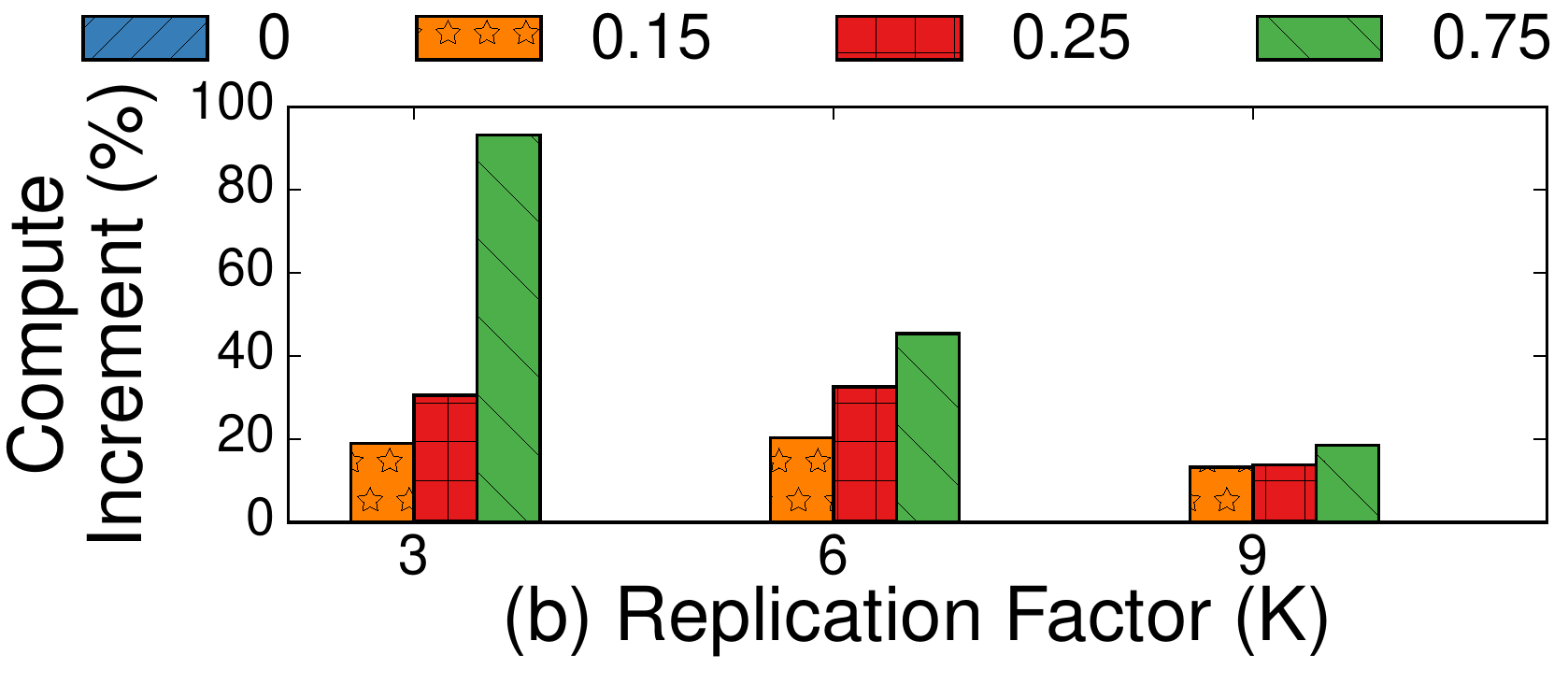}%
 		\label{sense-c}
 	}
 	\subfloat{%
 		\includegraphics[width=0.28\textwidth]{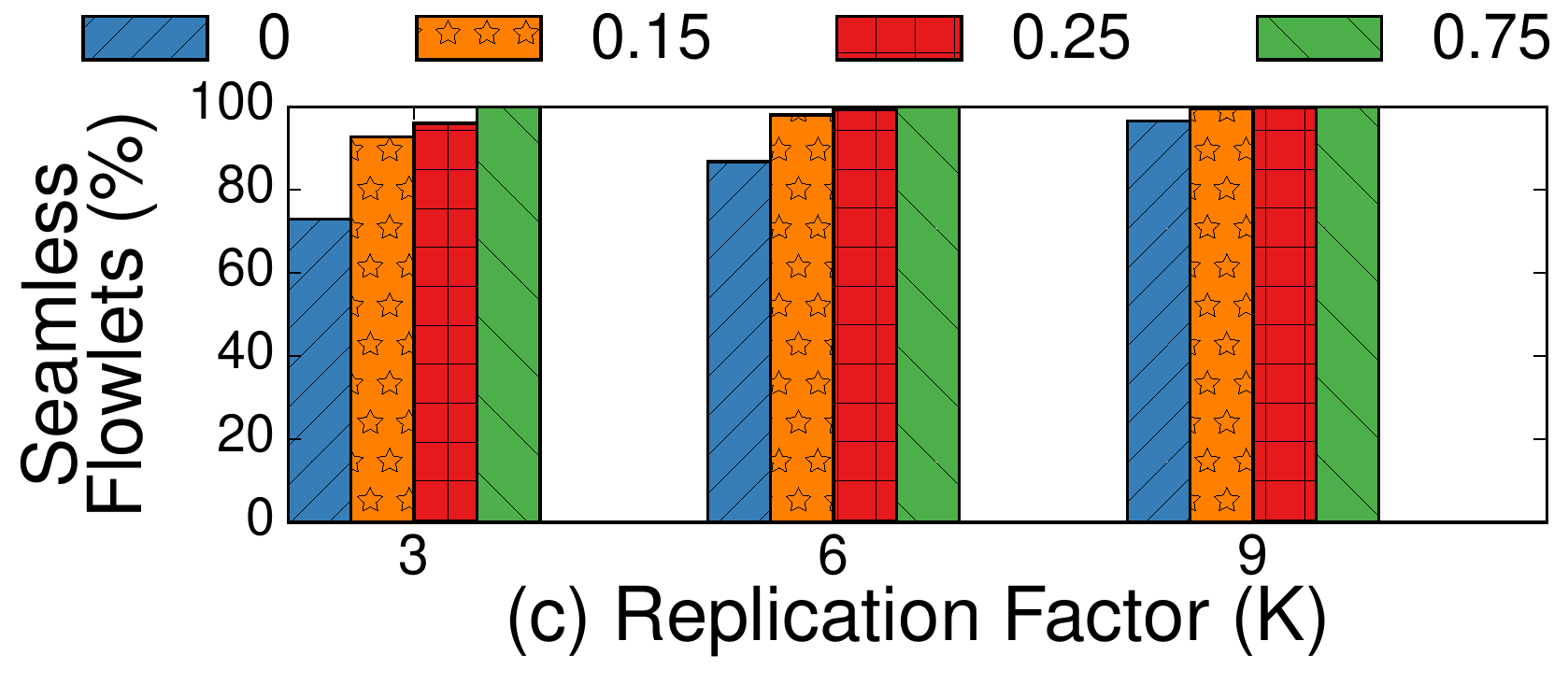}%
 		\label{sense-m}
 	}
  	\vspace*{-3mm}
 	\caption{ \footnotesize Impact of varying K and $\alpha$ on (a) packet processing time, (b) compute instances and (c) flowlets that can be processed seamlessly.}
 	\vspace*{-5mm}
 \end{figure*}

 \begin{figure}[h]
	\vspace*{-6mm}
	\centering 
	\subfloat{%
		\includegraphics[width=0.24\textwidth]{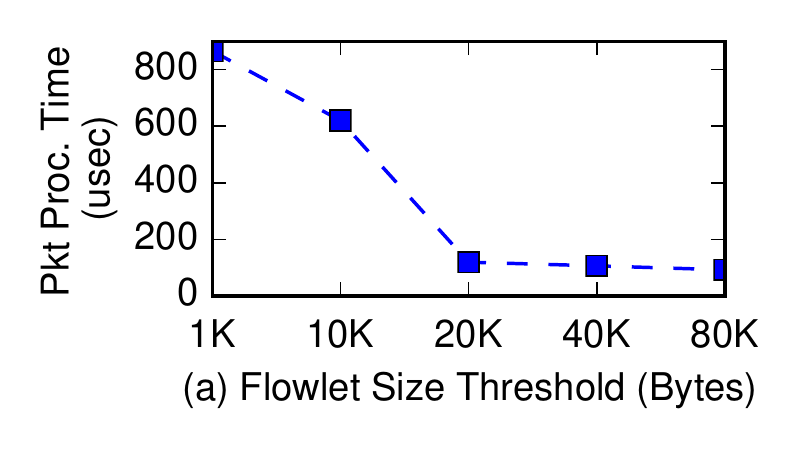}
		\label{sense-size-latency}
	}
	\subfloat{%
		\includegraphics[width=0.24\textwidth]{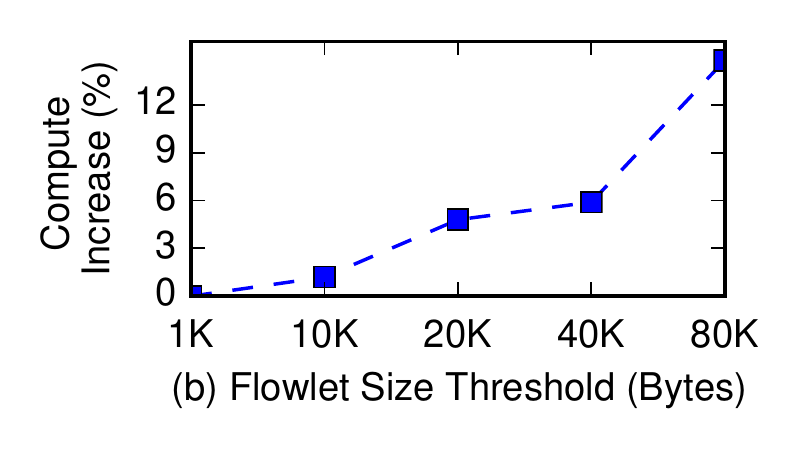}%
		\label{sense-size-compute}
	}
    \vspace*{-3mm}
	\caption{ \footnotesize Impact of size threshold on latency and compute provisioning}
\end{figure}

\begin{figure}[h]
	\vspace*{-8mm} 
	 \captionsetup[subfloat]{captionskip=-5pt}
	\centering
	\subfloat[][Independent Controllers]{
		\includegraphics[width=0.23\textwidth]{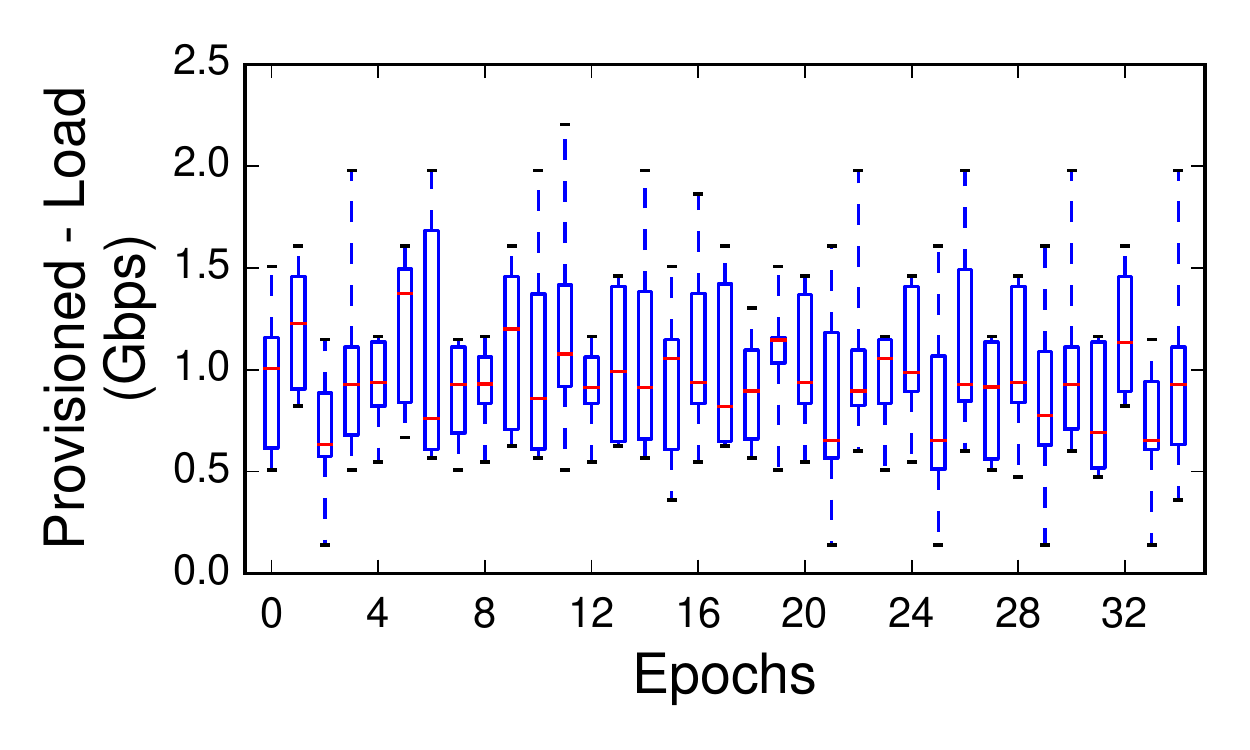}
		\label{parallel}
	}
	\subfloat[][SNF's Solution]{
		\includegraphics[width=0.23\textwidth]{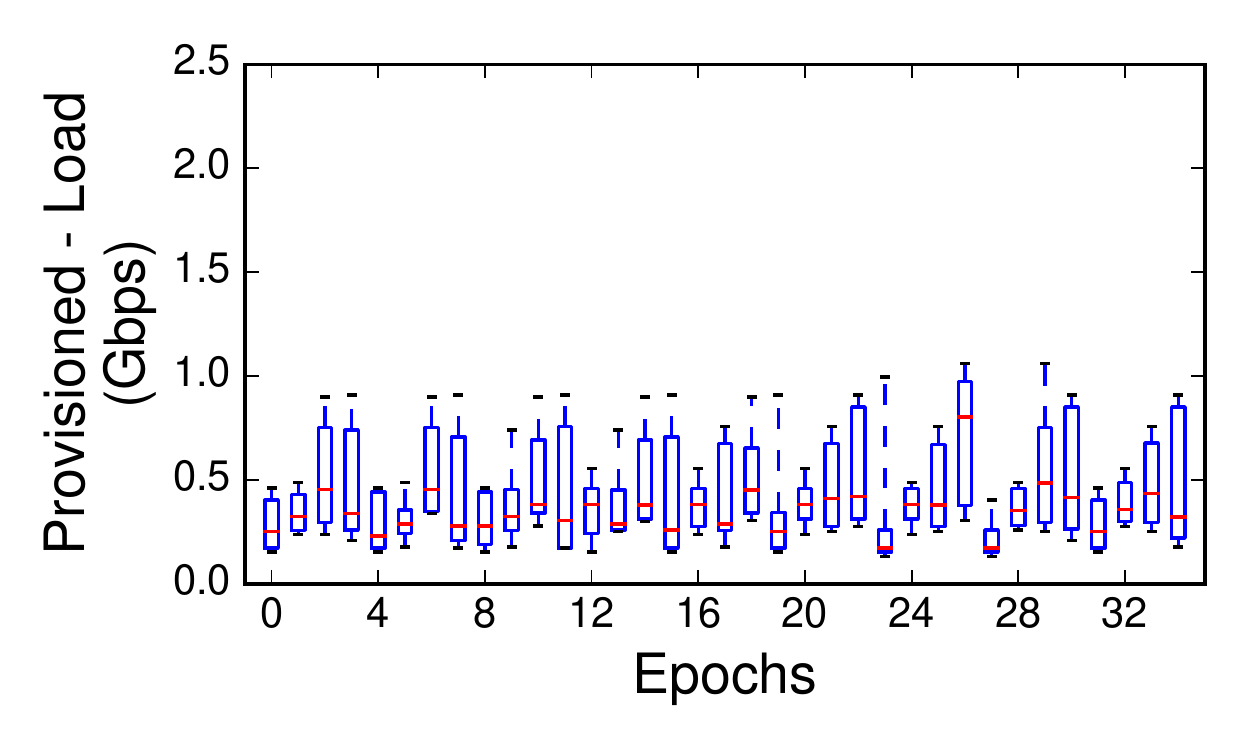}
		\label{shared}
	}
    \vspace*{-2mm}
	\caption{ \footnotesize Comparison of epoch-wise overprovisioning distribution by each controllers while using independent controllers and our approach}
	\vspace*{-2mm}
\end{figure} 

\subsection{Tackling Adversarial Flowlets}
\label{ss:feval}
In order to evaluate if \name can tackle adversarial flowlets we use three
synthetic workloads with varying frequency of such flowlets: we create these
flowlets every 100th, 50th and 10th flow by adding bursts of 20 packets
(\textasciitilde 1400 bytes) on average. Other
aspects of the experimental setup remain the same. For brevity, we only present results for
NAT below. 

Recall that \name should be able to mitigate the impacts of
adversarial flowlets as it uses a size threshold in addition to
inactivity timeout to detect flowlets. To study whether this helps, we
compare detecting flowlets using both the criteria (timeout + size)
with a baseline mode of using just the timeout (timeout) in terms of
the packet processing latencies.

We see in Fig.~\ref{high-load}, that it is indeed beneficial to use both
timeout and size in comparison to just using the timeout - from the least
aggressive to the most aggressive workload we observe that the median (tail)
latency reduces from 101$\mu$s (677.9$\mu$s) to 87$\mu$s (504.4$\mu$s),
201.2$\mu$ (1.5ms) to 93$\mu$s (702.4$\mu$s) and 752$\mu$ (5.2ms) to 102$\mu$s
(756.2$\mu$s). Using both helps \name bound the impact of adversarial flowlets
by starting a new flowlet as soon as the size threshold has been met, which
happens quickly for an adversarial flowlet and gives us the opportunity to
reallocate such flowlets (does not occur while using just timeout) leading to reduced packet processing latencies. 

\subsection{Fault Tolerance}
\label{ss:feval}

We study the performance of \name under failure recovery and compare it
against state of the art NF fault tolerance solutions - FTMB and CHC~\cite{ftmb,chc}. The main
metric of interest is the recovery time, i.e., the
amount of time it takes to ensure that a new NF unit is available with
up to date state. We fail a single NAT unit and measure
the recovery times for FTMB, CHC and \name at 50\% load. We assume that the failover compute unit is launched
immediately in all cases.  

In case of FTMB, the recovery time is 25.7ms (assuming that FTMB does
checkpointing every 50ms) and includes the time taken to load the
latest checkpoint as well as the time taken to process the packets
that need to be replayed to bring the new NAT instance up to date. CHC
under the same failure scenario takes 3.2ms during which the latest
state is fetched from the datastore and the in-transit packets are
replayed. On the other hand, in \name the recovery time is 140$\mu$s
which accounts for the amount of time taken to transfer the state from
the OL of the failed NAT unit to this newly launched NAT
unit. Unlike FTMB and CHC, given that
\name stores a copy of the latest state at the OL, it does not need to
replay packets during recovery leading to a faster recovery time.

\subsection{Multiple Controllers}
\label{ss:multiple-eval}
In order to evaluate the performance of \name at scale when using multiple
controllers we compare our approach of using a RM to the
baseline mode of operating the controllers independently. We use 10
controllers and the cumulative input load is on average 93.5 Gbps. We record
the per-controller provisioned capacity and actual load received and look at their difference (see Fig.~\ref{parallel}-\ref{shared}). The Y-axis value being 0 represents the ideal
case (provisioned capacity equals load), $>0$ indicates over provisioning
(reducing efficiency).

As seen in Figs.~\ref{parallel}-\ref{shared}, the amount of over
provisioning is minimal in case of \name as opposed to using
independent controllers. The reason being that in the baseline
mode there is more resource fragmentation due to controllers
provisioning compute units independently. With
\name, since the RM ``leases out'' capacity of
compute units, resource fragmentation is reduced as multiple
controllers can send traffic to the same shared compute unit (up to
their allocated share).

\subsection{Sensitivity Analysis}
\label{ss:sense-eval}

\noindent \textbf{Flowlet Inactivity Timeout (T): } Setting the flowlet
inactivity timeout plays a crucial role in \name as the value decides how
closely \name can adapt to traffic changes.
Additionally, it impacts the efficiency of proactive replication.


We consider two  timeout thresholds: $T =
100\mu$s and $T = 500\mu$s. In comparison to allocation at the per flow level, we see $5.18X$ and $4.66X$ more opportunities to do
work allocation when $T = 100\mu$s and $T = 500\mu$s, respectively.
While $T=100\mu$s clearly has benefits, we choose $T=500\mu$s in SNF. This improves the benefits of proactive replication, given that to replicate state of our NFs takes about 160$\mu$s.



\noindent \textbf{Flowlet Size Threshold (B): }
We consider multiple thresholds for B to study the impact on both performance and utilization for the NAT NF.
As the flowlet size increases, the processing
latencies improve (see Figs.~\ref{sense-size-latency}) primarily because the number of reactive state pulls decrease. 
On the other hand, this decrease in latency comes at the cost
of increased usage of compute units (see
Figs.~\ref{sense-size-compute}) due to poor packability of the larger work allocation units.

\noindent \textbf{Replication Factor (K) and Balancing Knob
  ($\alpha$):} 

Figs.~\ref{sense-l}-\ref{sense-m} show the impact of changing the
value of K and $\alpha$ while having a maximum of 15 compute units in
use. Here the NF capacity used is 500~Mbps.  For a given K, on increasing $\alpha$, the number of
flowlets that can be processed without state unavailability delays increases, as our scoring
metric gives more importance to units that have the state (\secref{sss:algo}). However,
this comes at the cost of using some amount of additional compute
units. Needless to say, the tail latencies improve as the value of
$\alpha$ increases as more flowlets are scheduled on units where their state is present (e.g., for $K=3$, the latency decreases from 107$\mu$s to 6.8$\mu$s when $\alpha$ changes from 0 to 0.75).
While for smaller values of $\alpha$, the latencies
are dominated by reactive state pulls, for larger values of $\alpha$ we
see that as K increases, the latency increases from 6.8$\mu$s to 15.2$\mu$s (when K changes from 3 to 9) reflecting the overhead 
involved in proactively replicating state.

\subsection{Overheads}
\label{ss:overheads}

\noindent \textbf{Work Allocation Overhead.} The \name controller calls into the
work allocation algorithm for every new flowlet. This adds an
additional latency of 1$\mu$s, but this is once per flowlet and hence
the cost is amortized across the packets of the flowlet.
 
\noindent \textbf{Proactive Replication.} In our current prototype, we proactively
replicate state to K compute units every 250us (half of the flowlet timeout).
For the said trace with K = 3, the proactive replication for NAT, LB, IDS, UDP Whitelister and QoS Traffic Policer 
uses up an additional bandwidth of 3.62 Mbps, 4.13 Mbps, 3.12 Mbps, 2.9 Mbps and 4.8 Mbps
respectively.

	\section{Other Related Work}

Some recent studies have show the benefits of using {\em existing} serverless
platforms in unmodified form for ``non-standard'' applications, e.g., scalable video
encoding~\cite{ex-camera}, and parallelized big data
computations~\cite{pywren,locus-nsdi19}. Other works instead focus on
improving key aspects of serverless computing, e.g., reducing
container start-up times~\cite{ed-atc18,boucher-atc18}, improved
system performance~\cite{akkus-atc18}, new storage
services~\cite{kilmovic-atc18,pocket-osdi18}, which proposed elastic ephemeral storage for serverless, and security~\cite{brenner-systor19}.  Our work falls into this second
category.

Our work adds to the long line of literature on network functions and
NFV. Improving performance in standalone software environments is the
goal of several papers~\cite{dpdk, netvm, bess, netbricks,
  nfp}. Several other systems tackle state management
issues~\cite{opennf, statelessNF,S6}. There has also been significant efforts in failure resiliency for NFV environments~\cite{ftmb,chc,reinforce}. 

	\section{Conclusions}
\label{sec:conc}

This paper shows the benefits of leveraging serverless computing for
streaming stateful applications, using the example of NFs. Our system
\name effectively tracks varying NF workload demands by elastically
allocation fine grained lambda-style compute resources, while ensuring
good efficiency.  \name decouples the unit at which functions operate
and the unit of serverless work allocation, using flowlets for the
latter. Additionally, we develop a peer-peer in-memory state storage
service that proactively replicates state during inter-flowlet gaps,
realizing ephemeral storage which is key to ensuring low per packet
processing latency.

	\label{EndOfPaper}
	{
	\raggedright
	\balance
	\bibliography{serverlessNF}
	\bibliographystyle{abbrv}
	}	
\end{document}